\documentclass[twocolumn,showpacs,prb,citeautoscript,amsmath,amssymb,floatfix,eqsecnum,superscriptaddress]{revtex4-2}
\usepackage{amsmath,amssymb,color}
\usepackage[dvipsnames]{xcolor}
\usepackage{bm}
\usepackage{graphicx}
\usepackage{psfrag}
\usepackage{bbold}
\usepackage{bbm}
\newcommand{\bd}{\bm}

\begin{document}

\title{Critical spin dynamics of Heisenberg ferromagnets revisited}

\author{Dmytro Tarasevych}

\affiliation{Institut f\"{u}r Theoretische Physik, Universit\"{a}t
  Frankfurt,  Max-von-Laue Strasse 1, 60438 Frankfurt, Germany}

\author{Peter Kopietz}
  
\affiliation{Institut f\"{u}r Theoretische Physik, Universit\"{a}t
  Frankfurt,  Max-von-Laue Strasse 1, 60438 Frankfurt, Germany}

\affiliation{Department of Physics and Astronomy,
	University of California, Irvine, California 92697, USA}

\date{January 4, 2022}

 \begin{abstract}

We calculate the dynamic structure factor $S (\boldsymbol{k},\omega)$ in the paramagnetic regime of quantum Heisenberg ferromagnets for temperatures $T$ close to the critical temperature $T_c$ using our recently developed functional renormalization group approach to quantum spin systems. In $d=3$ dimensions we find that for small momenta $\boldsymbol{k}$ and frequencies $\omega$ the dynamic structure factor assumes the scaling form $S(\boldsymbol{k},\omega) = (\tau T G (\boldsymbol{k})/\pi)\Phi (k\xi, \omega\tau)$, where $ G (\boldsymbol{k})$ is the static spin-spin correlation function, $\xi$ is the correlation length, and the characteristic time-scale $\tau$ is proportional to $\xi^{5/2}$. We explicitly calculate the dynamic scaling function $\Phi (x,y)$ and find satisfactory agreement with neutron scattering experiments probing the critical spin dynamics in EuO and EuS. Precisely at the critical point where $\xi = \infty$ our result for the dynamic structure factor can be written as $S (\boldsymbol{k},\omega) = (\pi\omega_k)^{-1} T_c G (\boldsymbol{k}) \Psi_c (\omega/\omega_k)$, where $\omega_k \propto k^{5/2}$. We find that $\Psi_c(\nu)$ vanishes as $\nu^{-13/5}$ for large $\nu$, and as $\nu^{3/5}$ for small $\nu$. While the large-frequency behavior of $\Psi_c (\nu)$ is consistent with calculations based on mode-coupling theory and with perturbative renormalization group calculations to second order in $\epsilon = 6-d$, our result for small frequencies disagrees with previous calculations. We argue that up until now neither experiments nor numerical simulations are sufficiently accurate to determine the low-frequency behavior of $\Psi_c (\nu)$. We also calculate the low-temperature behavior of $S ( \boldsymbol{k},\omega)$ in one- and two dimensional ferromagnets and find that it satisfies dynamic scaling with exponent $z=2$ and exhibits a pseudogap for small frequencies.

\end{abstract}

\maketitle

\section{Introduction}
 \label{sec:intro}

The spin dynamics of isotropic Heisenberg ferromagnets
for temperatures $T$ at and slightly above the critical temperature $T_c$
has been investigated for many decades both 
theoretically \cite{VanHove54,Kawasaki67,Wegner68,Halperin69,
Hohenberg77,Resibois70, Hubbard71,Ma75,Dohm76,Nolan77,
Borckmans77,Fogedby78,Bhattacharjee81,Folk85,Frey89,Frey94,Chen94,Tao05,Folk06} and 
experimentally via inelastic neutron scattering \cite{Passell76,Wicksted84,Mezei86,Boeni86,Boeni87,Boeni88}.
The central quantity of interest 
is the dynamic structure factor
 \begin{equation}
 S ( \bd{k} , \omega ) = \frac{1}{3N} \sum_{ij} \int_{- \infty}^{\infty} d t
 e^{ i \bd{k} \cdot ( \bd{R}_i - \bd{R}_j ) - i \omega t }
 \langle \bd{S}_i ( t ) \cdot \bd{S}_j ( 0 ) \rangle,
 \end{equation}
where the spin operators $\bd{S}_i$ are
localized at the sites $\bd{R}_i$
of a  Bravais lattice, the indices $i, j  = 1 , \ldots , N$ label the  lattice sites, and
the time evolution is in the Heisenberg picture.
In dimensions $d < 6$
the calculation of $S ( \bd{k} , \omega )$  in the vicinity
of the critical point of a Heisenberg ferromagnet 
is challenging 
because the dynamics is dominated by
non-Gaussian critical fluctuations \cite{Halperin69,Ma75,Dohm76,Nolan77}. 
In the spirit of the $\epsilon$-expansion of thermodynamic critical exponents
via renormalization group (RG) methods \cite{Wilson72},  
the critical dynamics has been investigated 
for small  $ \epsilon = 6-d$ by applying a dynamic renormalization group procedure to the relevant
stochastic equations of motion \cite{Ma75,Dohm76,Nolan77,Bhattacharjee81,Folk85}.  However, the problem is considerably more complicated than the
$\epsilon$-expansion of critical exponents, because
one is interested  in the spectral line-shape of $S ( \bd{k} , \omega )$ 
close to the critical point. Recall that
according to the dynamic scaling hypothesis \cite{Halperin69,Hohenberg77}
the dynamic structure factor
for small momenta and frequencies 
can be written in the scaling form
 \begin{equation}
 S ( \bd{k}  , \omega ) =   \frac{T\tau G ( \bd{k} )}{\pi} \Phi ( k \xi , \omega \tau ),
 \label{eq:Sscale1}
 \end{equation}
where $ G ( \bd{k} )$ is the static spin-spin correlation function, $\xi$ is the correlation length, $\tau$  is a characteristic time-scale, and
$\Phi ( x , y )$ is a dimensionless scaling function.
To obtain meaningful results for the scaling function $\Phi ( x , y )$ within a
perturbative RG approach,
some kind of interpolation procedure is necessary which resums the 
$\epsilon$-expansion \cite{Folk85}. 
Moreover, it is a priori not clear 
 whether  an extrapolation to the physically
relevant case $\epsilon = 3$ is possible. 
Nevertheless,
satisfactory agreement  between
RG calculations to first order in $\epsilon$
and neutron scattering experiments \cite{Wicksted84} 
have been reported \cite{Folk85}. 
However, according to  Ref.~[\onlinecite{Bhattacharjee81}] 
two-loop corrections corresponding to terms of order $\epsilon^2$
can significantly change the one-loop result for the spectral line-shape and it is not clear how even higher orders in $\epsilon$ would change the two-loop result for $d = 3$.

In principle, it should be possible to calculate the
scaling function $\Phi ( x , y )$ using modern 
functional renormalization group (FRG) methods \cite{Wetterich93,Berges02,Pawlowski07,Kopietz10,Metzner12,Dupuis21} which 
give
flow equations for momentum- and frequency dependent correlation functions and
do not rely on the small parameter
$\epsilon = 6-d$; the present work is a first step in this direction.
In fact, in our recent work on the spin dynamics of quantum paramagnets at infinite 
temperature \cite{Tarasevych21} we have used a variant of the FRG approach to quantum spin systems developed in Ref.~[\onlinecite{Krieg19}]
to derive an integral equation of the 
imaginary-frequency spin-spin correlation function $G ( \bd{k} , i \omega )$ of Heisenberg magnets in the paramagnetic phase. 
The latter is related to the  dynamic structure factor via the
fluctuation-dissipation theorem,
 \begin{equation}
 S ( \bd{k} , \omega ) = \frac{1}{ 1 - e^{ - \omega /T } } \frac{1}{\pi}
 {\rm Im} G ( \bd{k} , \omega + i 0 ).
 \label{eq:flucdis}
 \end{equation}
As discussed in Ref.~[\onlinecite{Tarasevych21}] and briefly summarized in the appendix,
in the paramagetic phase of a Heisenberg model
it is convenient to parametrize the imaginary-frequency spin-spin correlation function
in terms of
an energy scale $\Delta ( \bd{k} , i \omega )$ (which we have called 
dissipation energy in Ref.~[\onlinecite{Tarasevych21}]) as follows,
  \begin{equation}
 G ( \bd{k} , i \omega ) = G ( \bd{k} ) \frac{ \Delta ( \bd{k} , i  \omega ) }{ 
  \Delta ( \bd{k} , i \omega )  + | \omega |}.
 \label{eq:GMatsubara}
 \end{equation}
The dissipation energy satisfies the integral equation
 \begin{equation}
 \Delta ( \bd{k} , i \omega ) = \frac{1}{N} \sum_{\bd{q}} \frac{ V ( \bd{k} , \bd{q} ) }{
  \Delta ( \bd{q} , i \omega )  +   | \omega |  },
 \label{eq:Deltaint}
 \end{equation}
where the kernel is given by 
 \begin{eqnarray}
 V ( \bd{k} , \bd{q} ) & = & 
   T  G^{-1} ( \bd{k} ) G^{-1} ( \bd{q} ) 
\bigl[   {G} ( \bd{q} + \bd{k} ) Z ( \bd{q} , \bd{k} )
 \nonumber
 \\
 & &
 +   {G} ( \bd{q} -  \bd{k}  ) Z ( \bd{q} , - \bd{k} ) 
   -   2 {G} ( \bd{q} ) \bigr].
 \label{eq:Vdef}
 \end{eqnarray}
Here the vertex correction factor is
 \begin{equation}
 Z ( \bd{q} , \bd{k} ) = \left[
 1 + \frac{1}{2} [ J ( \bd{q} + \bd{k} ) - J ( \bd{q} ) ] G ( \bd{q} )
 \right]^2,
 \label{eq:Zdef}
 \end{equation} 
where  $J ( \bd{k} )$ is the Fourier transform of the exchange couplings.
In Ref.~[\onlinecite{Tarasevych21}] we have explicitly solved the integral equation 
(\ref{eq:Deltaint}) for various types of Heisenberg models at infinite temperature, where the static spin-spin correlation function
can be calculated systematically via an expansion in powers of $1/T$.
In this work  we will solve Eq.~(\ref{eq:Deltaint}) for an
isotropic  Heisenberg ferromagnet for temperatures close to the
critical temperature $T_c$, including  the critical point $T = T_c$.
Using the fluctuation-dissipation theorem (\ref{eq:flucdis}) we then obtain the
dynamic structure factor in the critical regime of a Heisenberg ferromagnet.
The necessary analytic continuation to real frequencies can be trivially performed because our integral equation (\ref{eq:Deltaint})  is local in frequency.

An alternative method to obtain the spin dynamics of Heisenberg magnets  is based on
the so-called mode-coupling theory \cite{Kawasaki66,Pomeau75,Goetze99,Das04}, where one 
derives an approximate integral equation for the Kubo relaxation 
function \cite{Mori65}
which is local in the time-domain and hence non-local in frequency space.
The structure of our integral equation is therefore very different from the
integral equation for the Kubo relaxation function of mode-coupling theory.
In fact, the locality in frequency considerably simplifies the solution of
 our integral equation (\ref{eq:Deltaint}).  
It is therefore not surprising that in some regimes our result for
$S ( \bd{k} , \omega )$ differ from the predictions of mode-coupling theory.
Because both mode-coupling theory and our approach based on truncated FRG
flow equations are approximate, it is a priori not clear which method gives more accurate results in the critical regime. In this work we will expicitly solve
Eq.~(\ref{eq:Deltaint}) in the critical regime of a Heisenberg ferromagnet and compare our results with perturbative RG calculations based on the
$\epsilon$-expansion, with mode-coupling theory, and with experiments.

\section{Dynamic structure factor in three dimensions}
 \label{sec:3d}

In this section we consider a spin-$S$ Heisenberg ferromagnet with short-ranged exchange on a three-dimensional Bravais lattice with cubic symmetry and lattice spacing $a$.  The Hamiltonian can be written as
 \begin{equation}
 {\cal{H}} = \frac{1}{2}\sum_{ij}J_{ij} \bd{S}_i \cdot \bd{S}_j.
 \end{equation}
 For a nearest neighbor coupling on a simple cubic lattice the corresponding Fourier transform of $J_{ij}$ reads
 \begin{equation}
 J ( \bd{k} ) = - 2 J [ \cos ( k_x a ) + \cos ( k_y a ) + \cos ( k_z a ) ],
 \end{equation}
 where $J > 0$ for a ferromagnet.
\subsection{Scaling regime above the critical temperature}

\begin{figure}[tb]
 \begin{center}
  \centering
\vspace{7mm}
\includegraphics[width=0.45\textwidth]{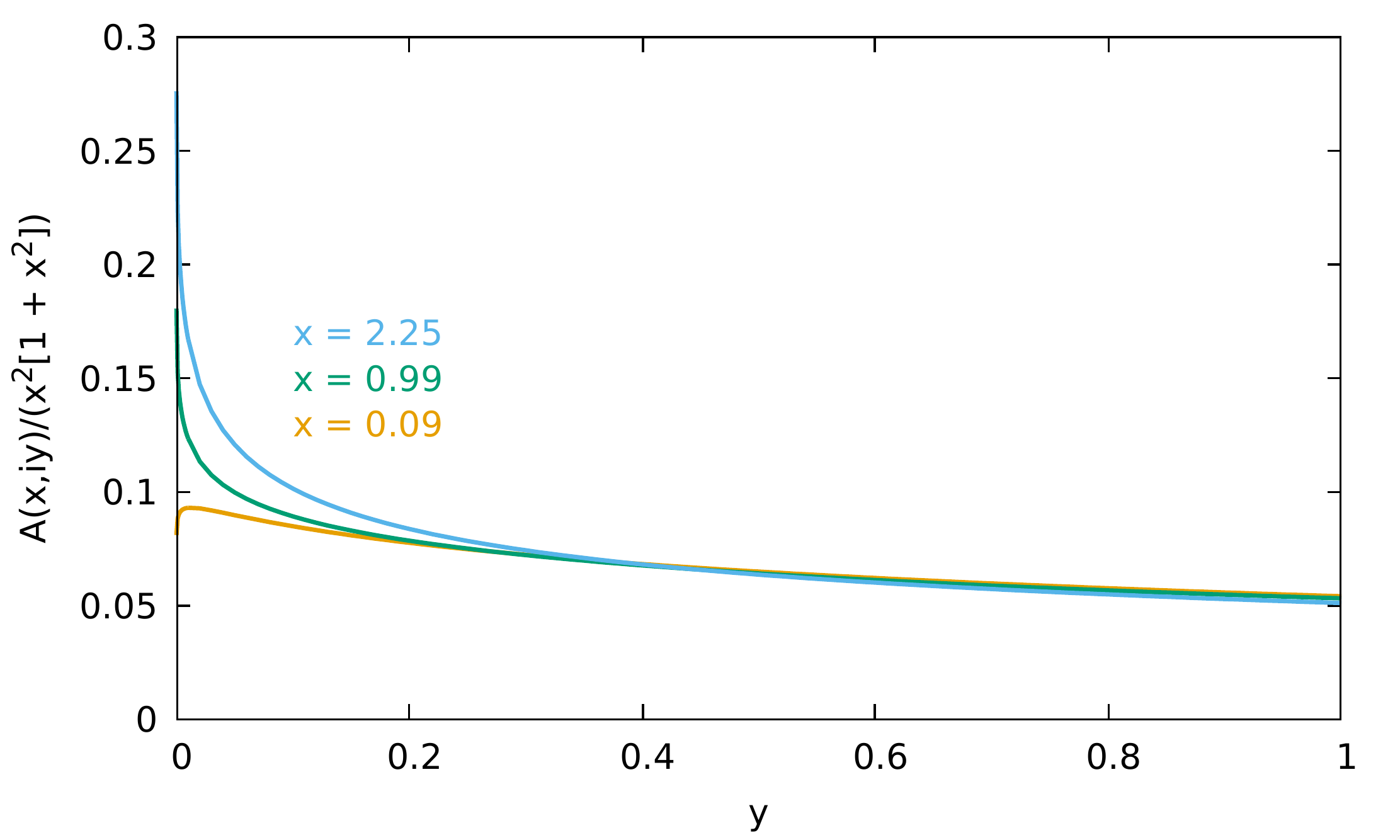}
   \end{center}
  \caption{Scaling function $A ( x , iy )$ of the dissipation energy $\Delta ( \bd{k} , i \omega )$  defined via Eq.~(\ref{eq:Deltascale}) 
 in $d=3$ obtained from the numerical solution
of the integral equation (\ref{eq:Aint1})  and divided by $x^{2}[1 + x^{2}]$. We plot $A(x,iy)/(x^2[1 + x^2])$, because this ratio depends only weakly on $x$.}
\label{fig:SAres}
\end{figure}

\begin{figure}[tb]
 \begin{center}
  \centering
\vspace{7mm}
\includegraphics[width=0.45\textwidth]{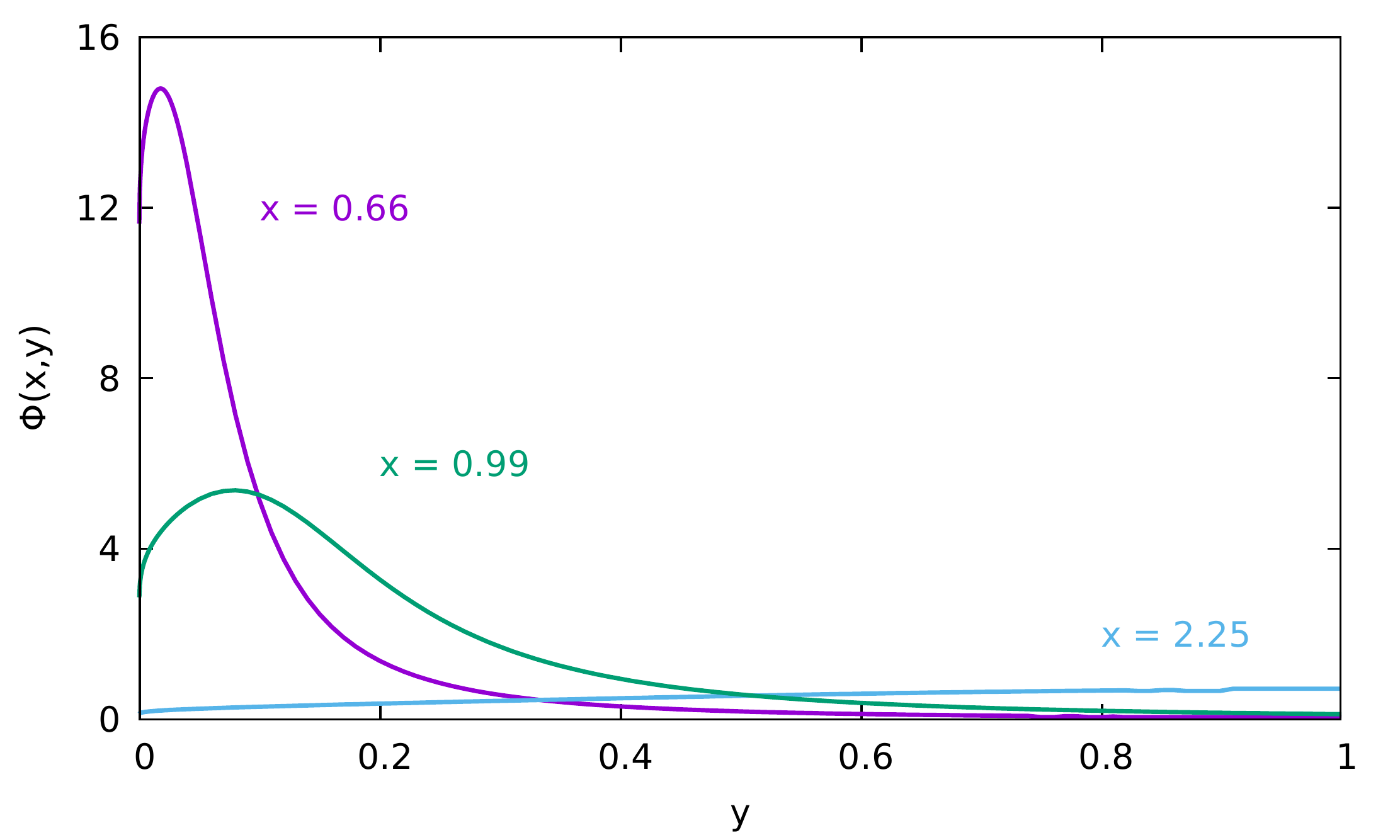}
   \end{center}
  \caption{Scaling function $\Phi ( x , y )$ of the dynamic structure factor
 $S ( \bd{k} , \omega )$
defined via Eq.~(\ref{eq:Sscale1}) in $d = 3$ as a function of the dimensionless frequency
$y = \omega \tau $ for different values of the dimensionless momentum
$x = k \xi$. 
}
\label{fig:Phiplot}
\end{figure}

\begin{figure}[tb]
 \begin{center}
  \centering
\vspace{7mm}
\includegraphics[width=0.45\textwidth]{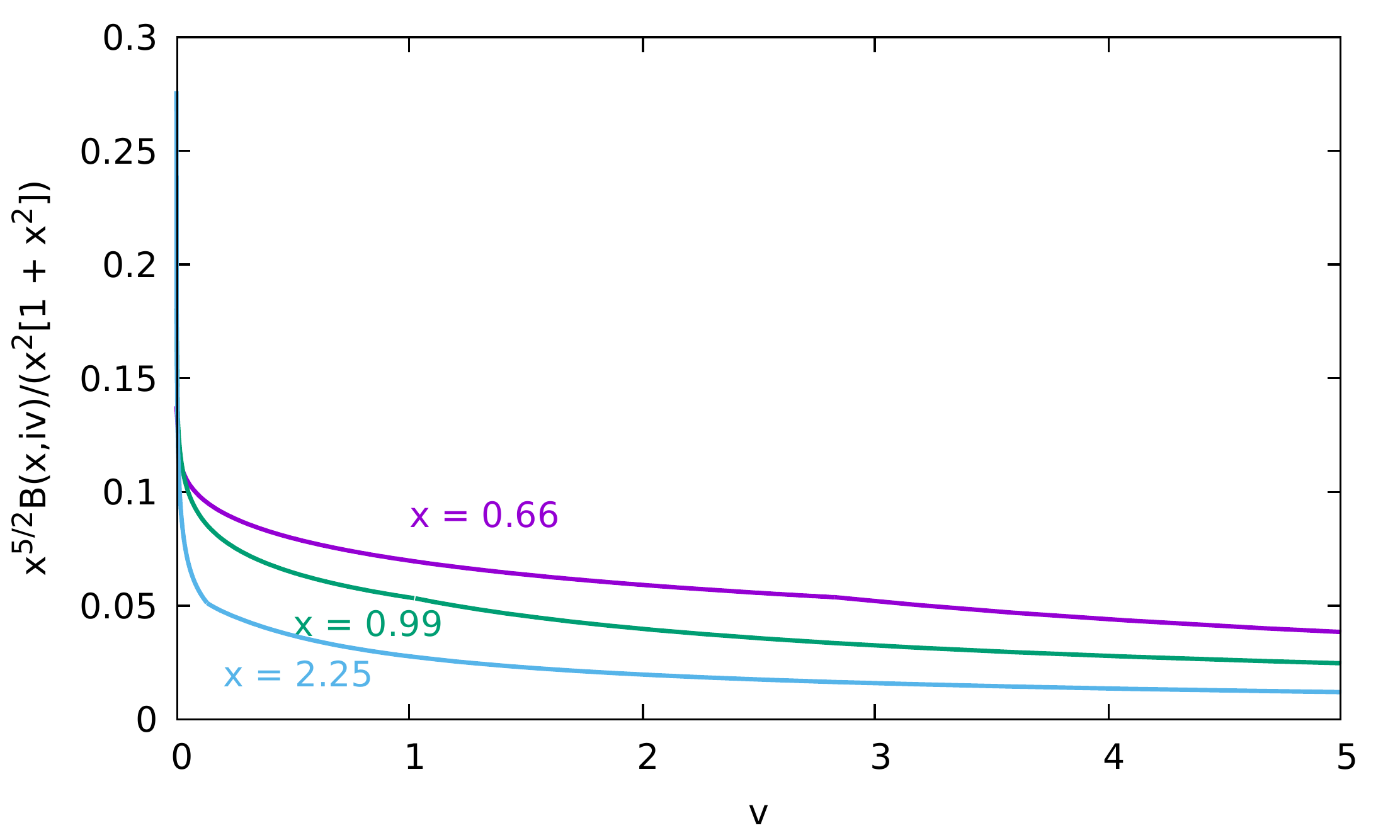}
   \end{center}
  \caption{Scaling function $B ( x , i \nu )$ of the dissipation energy $\Delta ( \bd{k} , i \omega )$
defined via Eq. \eqref{eq:Bdef} in $d=3$ obtained from the numerical solution
of the integral equation (\ref{eq:Bint1}).
We plot the ratio $B ( x , i \nu ) / ( x^{2-z} [ 1 + x^2 ])$ 
(which is the same as $A(x, i\nu x^{z})/(x^2[1+x^2])$)  because this quantity
exhibits again a relatively weak dependence on $x$.}
\label{fig:SBres}
\end{figure}

\begin{figure}[tb]
 \begin{center}
  \centering
\vspace{7mm}
\includegraphics[width=0.45\textwidth]{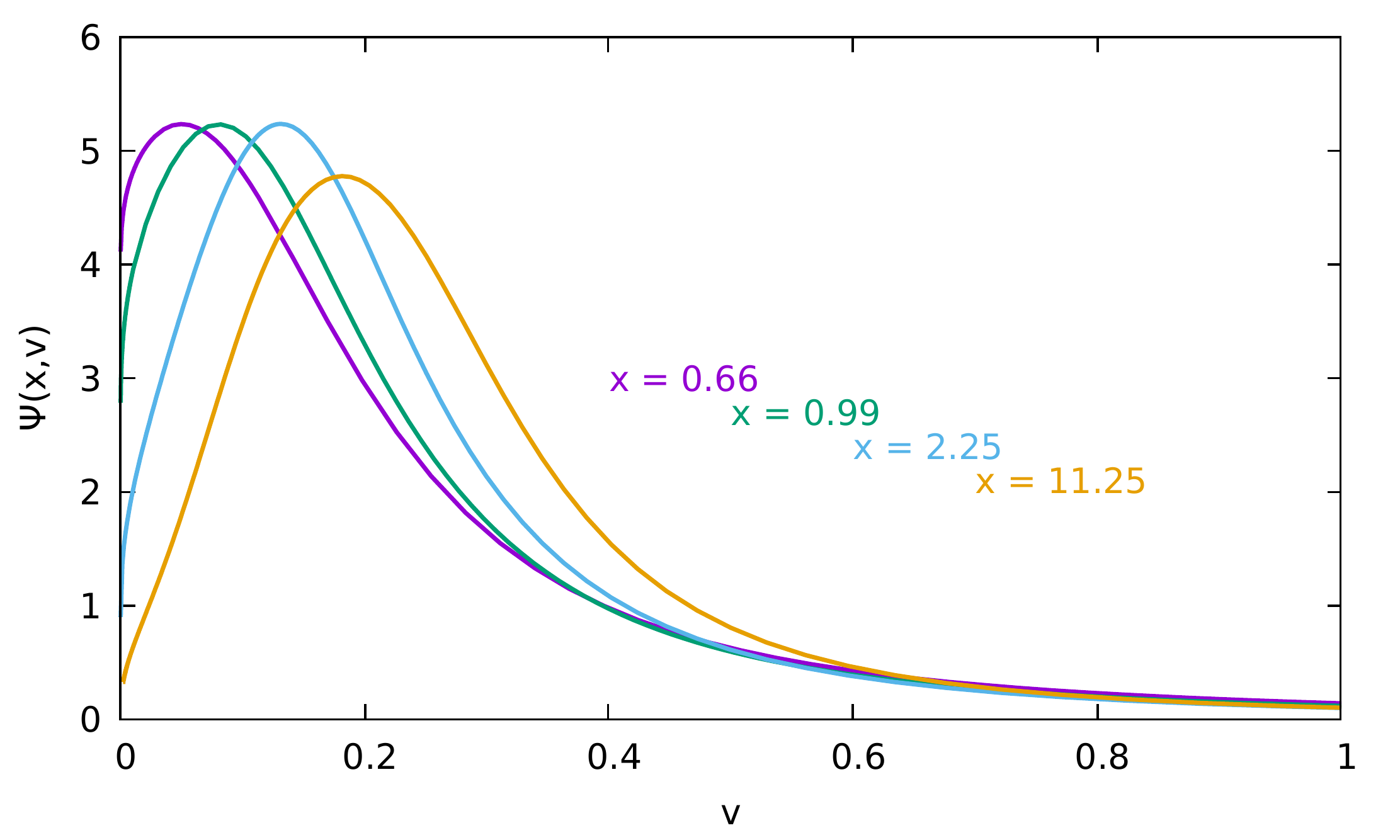}
   \end{center}
  \caption{Scaling function $\Psi ( x , \nu )$ of the dynamic structure factor
 $S ( \bd{k} , \omega )$
defined via Eq. \eqref{eq:PsiPhi} as a function of the ratio $\nu = \omega / \omega_k$
for different values of the dimensionless momentum
$x = k \xi$. 
}
\label{fig:Psiplot}
\end{figure}

To solve the integral equation (\ref{eq:Deltaint}) for the dissipation energy $\Delta ( \bd{k} , i \omega )$, we need the static spin-spin correlation function $G ( \bd{k} )$ which can be 
written as
 \begin{equation}
G(\bm{k}) =  \frac{1}{ J ( \bd{k} ) + \Sigma(\bm{k}) },
\label{eq:GkSigma}
\end{equation}
where $\Sigma ( \bd{k} )$ is the static irreducible self-energy.
With this definition  the kernel $V ( \bd{k} , \bd{q} )$ defined in Eq.~(\ref{eq:Vdef}) can also be written as 
 \begin{eqnarray}
 V ( \bd{k} , \bd{q} ) & = &  T G^{-1} ( \bd{k} ) G ( \bd{q} + \bd{k} )  \biggl[
 \frac{G ( \bd{q} )}{4}  [ J ( \bd{q} + \bd{k} ) - J ( \bd{q} )]^2
 \nonumber
 \\
 & & + \Sigma ( \bd{q}) - \Sigma ( \bd{q} + \bd{k} ) + ( \bd{k} \rightarrow - \bd{k} ) \biggr].
 \label{eq:kernel2}
 \end{eqnarray}  
In three dimensions we may
neglect the momentum dependence of the self-energy, which amounts to setting
 \begin{equation}
 \Sigma ( \bd{k} ) \approx \Sigma (0).
 \label{eq:sigmaapprox}
 \end{equation}
 This approximation is incompatible with
the finite value of  the anomalous dimension $\eta$ at the critical point.
However, in three dimensions the numerical value of 
$\eta \approx 0.027$ for the
Heisenberg universality class  \cite{Holm93} is rather small so that the finite value of $\eta$ has almost no
practical consequences. With the approximation (\ref{eq:sigmaapprox}) the kernel
$V ( \bd{k} , \bd{q} )$ in Eq.~(\ref{eq:kernel2}) simplifies to
\begin{eqnarray}
 V ( \bd{k} , \bd{q} ) & = &  \frac{T}{4} G^{-1} ( \bd{k} ) G ( \bd{q} + \bd{k} ) 
 G ( \bd{q} )  [ J ( \bd{q} + \bd{k} ) - J ( \bd{q} )]^2
 \nonumber
 \\
 & &  + ( \bd{k} \rightarrow - \bd{k} ) .
 \label{eq:kernel3}
 \end{eqnarray}  
Since we  are only interested in the dynamic structure factor for
small momenta $k a \ll 1$ and frequencies $\omega \ll J$, 
we may expand the Fourier transform of the exchange coupling $J ( \bd{k} )$ 
to quadratic order in $\bd{k}$,
\begin{equation}
J(\bm{k}) = J(0) +  J^{\prime \prime} (ka)^2  + \mathcal{O}(k^4),
\label{eq:Jexpansion}
\end{equation}
where for a ferromagnetic nearest-neighbor coupling on a cubic lattice
$J (0) = - 6 J$ and $ J^{\prime \prime}  = J $. The approximation (\ref{eq:Jexpansion})
is justified for $| T - T_c | \ll T_c$ in dimensions $d < 6$ because in this case the
leading singular contribution to the static susceptibility $ G ( \bd{k} )$ is dominated by
small momenta; the limits of momentum integrations can then be extended
to infinity as long as the relevant integrals are ultraviolet convergent.
In this approximation the static spin-spin correlation function $G ( \bd{k} )$ assumes for
$ka \ll 1$ the Ornstein-Zernike form
\begin{equation}
G(\bm{k}) = \frac{\chi}{ 1 + (k\xi)^2 },
\label{eq:oz_static}
\end{equation}
where 
 \begin{equation}
\chi = G(0) = \frac{1}{ J (0) + \Sigma (0 )}
 \end{equation}
is the  uniform susceptibility and the square of the correlation length $\xi$ is given by
 \begin{equation}
 \xi^2 =  \rho_0 \chi,
 \end{equation}
with the bare spin stiffness 
 \begin{equation}
 \rho_0 = J^{\prime \prime} a^2.
 \end{equation}
Note that at the critical temperature $J (0) +  \Sigma (0) =0$ and hence $\xi = \infty$.
If we approximate the static self-energy by its leading  high-temperature expansion,
$\Sigma (0 ) \approx 3 T /(S(S+1))$, we obtain the usual mean-field estimate for the
critical temperature, $T_c \approx | J (0 ) | S (S+1)/3$.
Substituting the Ornstein-Zernike form (\ref{eq:oz_static}) for the static spin-spin correlation function into our approximate  expression (\ref{eq:kernel3}) for the integral kernel
$V ( \bd{k} , \bd{q} )$ we obtain
 \begin{equation}
 V ( \bd{k} , \bd{q} ) = \frac{T \rho_0  }{4} \frac{[1 + (k \xi )^2] 
[ k^2 + 2 \bd{k} \cdot \bd{q} ]^2 \xi^2 }{
 [ 1 + ( q \xi )^2 ][  1 + ( \bd{k} + \bd{q} )^2 \xi^2 ] }
 + ( \bd{k} \rightarrow - \bd{k} ).
 \label{eq:Vkernel0}
 \end{equation}
In $d$ dimensions our integral equation (\ref{eq:Deltaint})
for the dissipation energy then reduces to
 \begin{eqnarray}
  \Delta ( \bd{k} , i \omega ) & = & \frac{Tv\rho_0}{2} a^d \int \frac{ d^d q}{ ( 2 \pi )^d}
  \frac{[ 1 + (k \xi )^2] [ k^2 + 2 \bd{k} \cdot \bd{q} ]^2 \xi^2    }{
 [ 1 + ( q \xi )^2 ][ 1 + ( \bd{k} + \bd{q} )^2 \xi^2 ]}
 \nonumber
 \\
 & & \hspace{20mm} \times
 \frac{ 1 }{ \Delta ( \bd{q} , i \omega ) + | \omega | },
 \label{eq:Deltapara}
 \end{eqnarray}
 where $v$ is the ratio of the volume of a single primitive unit cell to the volume  $a^d$ of  the conventional unit cell.
Introducing the characteristic time scale
 \begin{equation}
 \tau = \sqrt{\frac{2}{TvJ^{\prime\prime}}}( \xi/ a )^{ z},
 \label{eq:taudef}
 \end{equation}
where 
 $
 z = 1 + d/2
 $
 is the dynamic exponent,  
the dissipation energy can be written in the scaling form
 \begin{equation}
  \Delta ( \bd{k} , i \omega )   = \tau^{-1} A ( k \xi, i \omega \tau ),
 \label{eq:Deltascale}
 \end{equation}
where the dimensionless scaling function $A ( x, i y )$ satisfies the integral equation
 \begin{eqnarray}
 A ( x , i y ) & = & [1 + x^2]\int \frac{ d^d r}{ ( 2 \pi )^d}  
 \frac{  ( x^2 + 2 \bd{x}
 \cdot \bd{r} )^2}{ ( 1 + r^{ 2}) [ 1 + ( \bd{x} + \bd{r} )^2 ]}
 \nonumber
 \\
 & & \hspace{15mm} \times
 \frac{1}{A (r , i y ) + | y |},
 \label{eq:integralg}
 \end{eqnarray}
which can also be written as
\begin{eqnarray}
 A ( x , i y ) & = & [1 + x^2] \int \frac{ d^d r}{ ( 2 \pi )^d}  
 \left[ \frac{x^2}{ 1 + r^2} - 
 \frac{   x^2 + 2 \bd{x}
 \cdot \bd{r} }{  1 + ( \bd{x} + \bd{r} )^2 } \right]
 \nonumber
 \\
 & & \hspace{15mm} \times
 \frac{1}{A (r , i y ) + | y |}.
 \label{eq:integralg2}
 \end{eqnarray}
The corresponding scaling form of the imaginary-frequency spin-spin correlation function is
 \begin{equation}
 G ( \bd{k} , i \omega ) = G ( \bd {k} ) \frac{ A ( k \xi , i \omega \tau )}{
 A ( k \xi , i \omega \tau ) + | \omega | \tau }.
 \end{equation}
Substituting this into the fluctuation dissipation theorem (\ref{eq:flucdis})
we conclude that for small frequencies, where
 \begin{equation}
 \frac{1}{ 1 - e^{ - \omega / T }} \approx \frac{T}{\omega },
 \label{eq:classflucdiss}
 \end{equation}
the dynamic structure factor can be written in the scaling form
 \begin{equation}
 S ( \bd{k} , \omega ) = \frac{T G ( \bd{k} )}{\pi \omega}
 {\rm Im} \left[
 \frac{ A ( k \xi , \omega \tau + i 0)}{ A ( k \xi , \omega \tau + i 0 ) - i \omega \tau }
 \right].
 \end{equation}
Obviously, this has the scaling form anticipated in Eq.~(\ref{eq:Sscale1}) with scaling function
 \begin{equation}
 \Phi ( x , y ) = \frac{1}{y } {\rm Im} \left[ \frac{ A ( x , y + i 0 )}{
 A ( x , y + i 0 ) - i y } \right] .
 \label{eq:phiscale}
 \end{equation}
To explicitly calculate the scaling functions in three dimensions, we note that in this case
the angular integration in Eq.~(\ref{eq:integralg}) can be carried out exactly so that 
$A ( x , iy )$ can be calculated by solving the one-dimensional integral equation
  \begin{eqnarray}
  A ( x , i y ) & = & \frac{1+ x^2}{2 \pi^2} \int_0^{\infty} dr \frac{r^2}{ A ( r , iy ) + | y | }
 \biggl[  \frac{ x^2 }{ 1 + r^2 }  
 \nonumber
 \\
 &   & + \frac{ 1+ r^2}{4 x r } \ln \left( \frac{ 1 + | r +x |^2}{1 + | r-x |^2 } \right) -1
   \biggr].
 \label{eq:Aint1}
 \end{eqnarray}
In Fig.~\ref{fig:SAres} we present our numerical results for $A ( x , i y )$ as a function of $y$ for
different values of $x$.
The corresponding scaling function $\Phi ( x , y )$ of the dynamic structure factor
given in Eq.~(\ref{eq:phiscale}) is shown in Fig.~\ref{fig:Phiplot}.
For large values of $x = k\xi$ the dynamic structure factor exhibits a peak with finite width and dispersion $\tilde \omega_k \propto k^{5/2}$.
To see this more clearly, it is convenient to express the scaling functions in terms of
the ratio
 \begin{equation}
 \nu = \frac{y}{ x^z} = \frac{ \omega \tau }{ ( k \xi )^z } = \frac{\omega}{\omega_k },
 \end{equation}
with the characteristic frequency 
 \begin{equation}
 \omega_k = \frac{ (k \xi )^z}{\tau} = \omega_\ast  ( k a )^z,
 \end{equation}
and the non-universal energy scale $\omega_{\ast}$ defined as
 \begin{equation}
  \omega_\ast = \sqrt{\frac{Tv J^{\prime \prime}}{2} }.
  \label{eq:omega_ast}
 \end{equation}
Eliminating $y = \omega \tau $ in favour of $\nu = \omega / \omega_k$ 
as independent variable, we can write the scaling forms of
the dissipation energy and the dynamic structure factor as follows,
 \begin{eqnarray}
 \Delta ( \bd{k} , i \omega ) & = &  \omega_k B ( k \xi, i \omega / \omega_k ),
 \label{eq:DeltaB}
 \\
 S ( \bd{k} , \omega ) & = & \frac{ T G ( \bd{k} )}{\pi\omega_k} \Psi ( k \xi , \omega / \omega_k ).
 \label{eq:SPsi}
 \end{eqnarray}
Comparing these definitions with Eqs.~(\ref{eq:Deltascale}) and (\ref{eq:Sscale1})
we see that
 \begin{eqnarray}
 B ( x , i \nu ) & = &  x^{-z} A ( x , i \nu x^z ),
 \label{eq:Bdef}
 \\
 \Psi ( x , \nu )  & = & x^z \Phi ( x , \nu x^z ).
 \label{eq:PsiPhi}
 \end{eqnarray}
The relation~(\ref{eq:phiscale}) implies that the scaling function $\Psi ( x , \nu )$ can be expressed in terms of $B ( x , \nu + i 0 )$ as follows,
 \begin{equation}
 \Psi ( x , \nu ) = \frac{1}{\nu} {\rm Im}
 \left[ \frac{ B ( x , \nu + i 0 )}{ B ( x , \nu + i 0) - i \nu } \right].
 \label{eq:PsiBscale}
 \end{equation}
Substituting $r = x \rho$ in Eq.~(\ref{eq:Aint1}) we find that
in three dimensions, where $z = 5/2$ the scaling function $B ( x , i \nu )$ satisfies the integral equation
 \begin{eqnarray}
 & & B ( x , i \nu )  =  \frac{1 + x^2}{2\pi^2 x^2}
 \int_0^{\infty} d \rho \frac{ \rho^2}{ \rho^{z} B ( x \rho , i \nu / \rho^{z} ) + | \nu | }
 \nonumber
 \\
 &  & \times \left[
 \frac{ x^2}{ 1 + x^2 \rho^2} + \frac{ 1 + x^2 \rho^2}{ 4 x^2 \rho}
 \ln \left( \frac{ 1 + x^2 | \rho +1 |^2}{ 1 + x^2 | \rho -1 |^2 }\right)
 -1 \right].
 \nonumber
 \\
 & &
 \label{eq:Bint1}
 \end{eqnarray}
Numerical results for the scaling functions $B ( x , i \nu )$ and $\Psi ( x , \nu )$
in three dimensions are shown in Figs.~\ref{fig:SBres} and \ref{fig:Psiplot}. 
Note that the peak position of $\Psi ( x , \nu )$ approaches a finite value for $x \gg 1$,
implying that $\tilde \omega_k  \propto k^{5/2}$ indeed can be identified  with the dispersion of an overdamped critical mode.

\subsection{Spin diffusion close to the critical point}

For hydrodynamic frequencies $ | y |  \ll 1$ we  may approximate 
$A ( x, iy ) \approx A ( x , 0 )$ in Eq.~(\ref{eq:Aint1}) which then reduces to
 \begin{eqnarray}
  A ( x , 0 ) & = & \frac{1+ x^2}{ 2 \pi^2} \int_0^{\infty} dr \frac{r^2}{ A ( r , 0 ) }
 \biggl[  \frac{ x^2 }{ 1 + r^2 }  
 \nonumber
 \\
 &   & + \frac{ 1+ r^2}{4 x r } \ln \left( \frac{ 1 + | r +x |^2}{1 + | r-x |^2 } \right) -1
   \biggr].
 \label{eq:Aint2}
 \end{eqnarray}
Assuming in addition $ | x | \ll 1$ corresponding to hydrodynamic momenta
we obtain
 \begin{equation}
 A ( x , 0 ) = A_2 x^2 + {\cal{O}} ( x^4 ),
 \end{equation}
where from the numerical solution of the integral equation 
(\ref{eq:Aint2}) we find $A_2 \approx 0.078$ in three dimensions.
We conclude that to leading order in $k \xi \ll 1$ and $ \omega \tau \ll 1$ 
the dissipation function is given by
 \begin{equation}
 \Delta ( \bd{k} , i \omega ) \approx  A_2 \frac{ ( k \xi )^2}{\tau }  = {\cal{D}} k^2,
 \label{eq:Deltadif}
 \end{equation}
with spin-diffusion coefficient
 \begin{equation}
 {\cal{D}} = A_2   \frac{\xi^2 }{ \tau}  = A_2 \sqrt{\frac{T vJ^{\prime \prime}}{2}}\frac{ a^2}{
    ( \xi / a  )^{\frac{d -2}{2} } }.
 \end{equation}
In particular, for $d=3$ we find that
the spin-diffusion coefficient vanishes for $T \rightarrow T_c$  
as $\xi^{-1/2} \propto \chi^{-1/4}$, 
in agreement with  Ref.~[\onlinecite{Kawasaki67}] and with the prediction of
dynamic scaling in the hydrodynamic regime \cite{Halperin69}.
Note that the older theory by Van Hove \cite{VanHove54} 
predicts  $\mathcal{D} \propto \xi^{-2} \propto \chi^{-1}$, which 
corresponds to the dynamic exponent $z = 4$;   it turns out that for isotropic ferromagnets
the Van Hove theory is only valid for dimensions   above the upper critical dimension $d_c = 6$.

Within the approximation (\ref{eq:Deltadif}) the dynamic structure factor exhibits a diffusive zero-energy peak
 \begin{eqnarray}
 S(  \bd{k} , \omega ) & = &  \frac{1}{ 1 - e^{ - \omega /T }} \frac{1}{\pi} {\rm Im} G ( \bd{k} , \omega + i 0 )
 \nonumber
 \\
 & = &  \frac{G ( \bd{k} ) }{ 1 - e^{ -  \omega/T }} \frac{1 }{\pi} {\rm Im} \frac{ {\cal{D}} k^2 }{ {\cal{D}} k^2 - i \omega }
 \nonumber
 \\
 & \approx &  \frac{T \chi  }{\pi} \frac{ {\cal{D}} k^2 }{ ( {\cal{D}} k^2)^2 + \omega^2 }.
 \label{eq:dynfinal}
 \end{eqnarray}
Corrections to hydrodynamics can be obtained by retaining the leading $y$-dependence 
of the scaling function $A ( x , iy )$ of the dissipation energy. For $d=3$ we obtain
\begin{equation}
A ( x , iy )  = A ( x, 0 )  + A_1 (x )  | y |^{1/2} + \mathcal{O}(y),
\label{eq:expanded_diss_energy}
\end{equation}
which can be seen by writing under the integral in Eq.~\eqref{eq:Aint1}
\begin{equation}
\frac{1}{ A(r, iy ) + | y |} = \frac{1}{ A ( r , iy)}\left[ 1 -  \frac{|y |}{A(r, i y ) + | y |} \right],
\end{equation}
and noting that for $ | y |  \ll 1$ 
the second term on the right-hand side becomes singular 
so that the integral can be restricted to the regime 
$r \lesssim \mathcal{O}(|y|^{1/2})$.
Note that  for sufficiently large momenta $x \gtrsim \mathcal{O}(1)$ the narrowing of the scaling function $\Psi ( x, \nu )$ of the dynamic structure factor shown in Fig.~\ref{fig:Psiplot}
for $\nu \ll x^{-z}$, which is accompanied by  a finite minimum at $\nu = 0$ and  symmetric maxima at finite $\nu$, can be explained in terms of the  non-analytic correction in Eq.~(\ref{eq:expanded_diss_energy})
with a negative $A_1 ( x )$ in this regime. In the opposite limit of small $x$
the function $\Psi ( x, \nu )$ exhibits only a single maximum at vanishing frequency which is related to the fact that
in this regime  the sign of $A_1 ( x)$ is positive. The latter behavior  of $A_1(x)$ is illustrated in Fig. \ref{fig:SAres} for the case $x = 0.09$.

According to Fogedby and Young \cite{Fogedby78}, at high temperatures (i.e., in the 
non-critical regime) the leading non-analytic frequency-dependence of the
generalized diffusion coefficient ${\cal{D}} ( \omega )$
of a paramagnetic spin system is in three dimensions proportional to $ \omega^{3/2}$.
Our result (\ref{eq:expanded_diss_energy}) in the critical regime implies a larger
$\omega^{1/2}$ correction which is nevertheless negligible
 in the {\it{strict hydrodynamic}} limit \cite{Borckmans77,Fogedby78, Pomeau75} given by the prescription
 $x = k \xi \rightarrow 0$ and $y = \omega \tau  \propto  x^2 \rightarrow 0$,
implying that the leading correction to the dissipation energy $\Delta(  {\bd{k}} , \omega \propto k^2)$ scales as $k^3$. In the momentum-time domain this limit corresponds to arbitrarily small momenta and long times $t$ that are constrained by constant $\mathcal{D}k^2 t $. The Fourier transform
$S(\bm{k},t)$ of the dynamic structure factor to the time-domain is then
dominated by the diffusion pole in Eq.~(\ref{eq:dynfinal}), 
\begin{equation}
S(\bm{k},t) \propto  \int_{-\infty}^{\infty}  d \omega
\frac{ {  e^{ i \omega t }      \cal{D}} k^2 }{ ( {\cal{D}} k^2)^2 + \omega^2 }  \propto
 e^{-\mathcal{D}k^2 t}.
\end{equation}
On the other hand, if we fix the momentum $\bd{k}$ and consider the limit of arbitrary small frequencies or long times,  the non-analytic term of order $ |y|^{1/2}$ in Eq.~(\ref{eq:expanded_diss_energy}) dominates
the asymptotics because it implies a branch point at vanishing frequency.
In the time domain the presence of this term produces a purely algebraic contribution 
 $\sim (\mathcal{D}k^2t)^{-3/2}$, which decays much slower 
than the exponential generated by the diffusion pole.  
Given the fact that the non-analytic term $A_1 ( x ) | y|^{1/2}$ exists for arbitrary $x = k \xi$,
the general long-time asymptotics in the time-domain is proportional to
$ [ A(x,0 ) t]^{-3/2}$. From this we conclude that in three dimensions 
the on-site autocorrelation function  decays for $ t \rightarrow \infty$ as
\begin{equation}
\frac{\langle {\bd{S}}_i (0) \cdot \bd{S}_i ( t ) \rangle}{3}  \equiv \frac{1}{N} \sum_{\bd{k}}   S(\bm{k},t)  \sim C_3  t^{-3/2},
 \label{eq:C3def}
\end{equation}
where the value of $C_3$ is not only determined by the diffusion pole but also by the
leading non-analytic correction in Eq.~(\ref{eq:expanded_diss_energy}) \cite{footnoteCor}.

Non-hydrodynamic corrections to diffusion have been  discussed 
previously in the literature \cite{Borckmans77,Fogedby78, Pomeau75}. 
However, in these works the non-analytic terms appear as functions of 
 $\mathcal{D}^{\prime} k^2 -i\omega$ with ${\cal{D}}^{\prime} < {\cal{D}}$, so that the
branch points occur at finite frequencies for $\bd{k} \neq 0$.
As a result, the branch-cut contribution to  $S(\bm{k},t)$ contains an additional exponential modulation on top of the power-law tails, which is absent in our approach.

 \subsection{Scaling at the critical point}
\begin{figure}[tb]
 \begin{center}
  \centering
\vspace{7mm}
\includegraphics[width=0.45\textwidth]{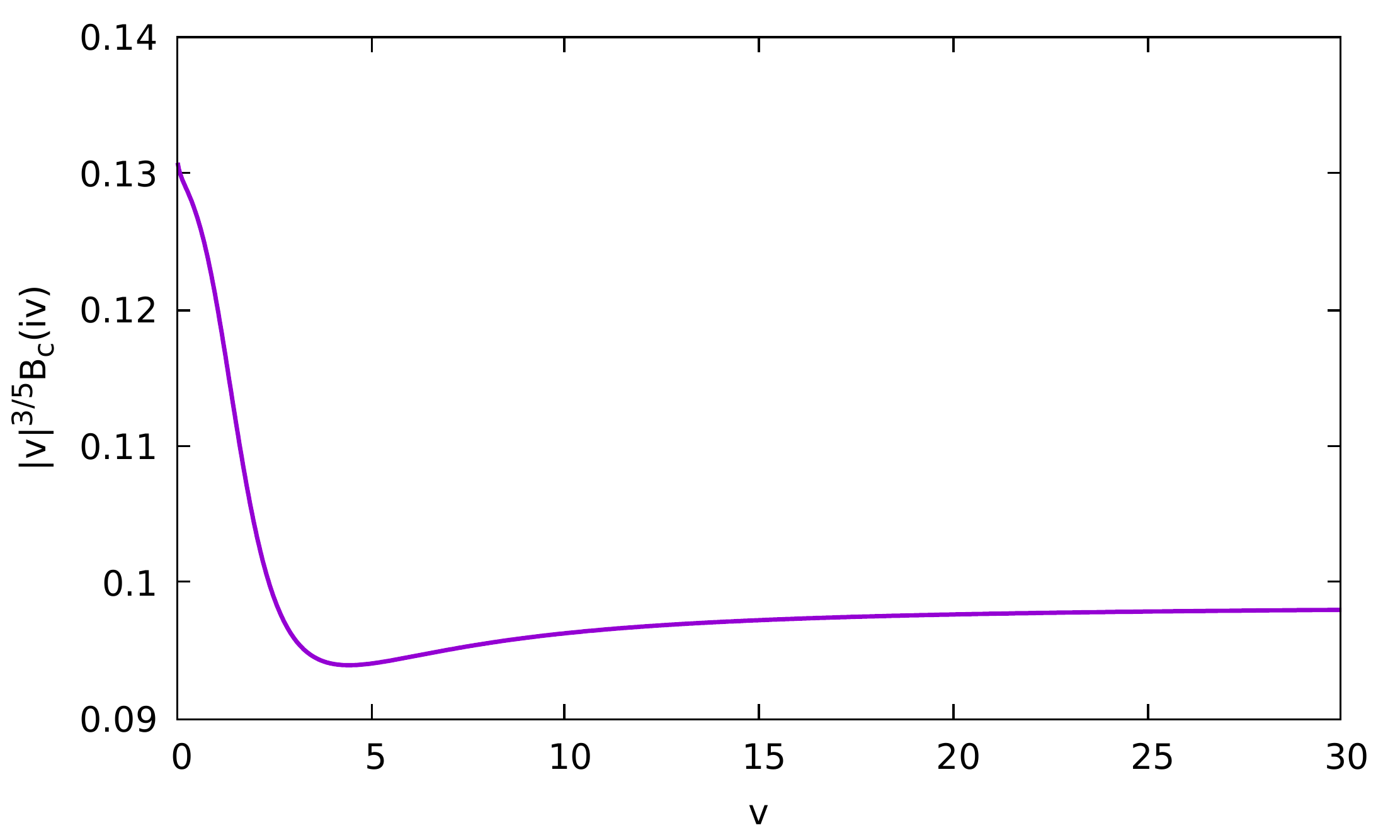}
   \end{center}
  \caption{Critical scaling function $B_c (i \nu )$ of the dissipation energy $\Delta ( \bd{k} , i \omega )$ defined via Eq.~(\ref{eq:Bcdef}) for $d=3$. Given the fact that 
$B_c ( i \nu )$ is proportional to $ | \nu |^{-3/5}$ for large and for small $ | \nu |$, we plot
$| \nu |^{3/5} B_c ( i \nu )$.
}
\label{fig:Bcrit}
\end{figure}

\begin{figure}[tb]
 \begin{center}
  \centering
\vspace{7mm}
  \includegraphics[width=0.45\textwidth]{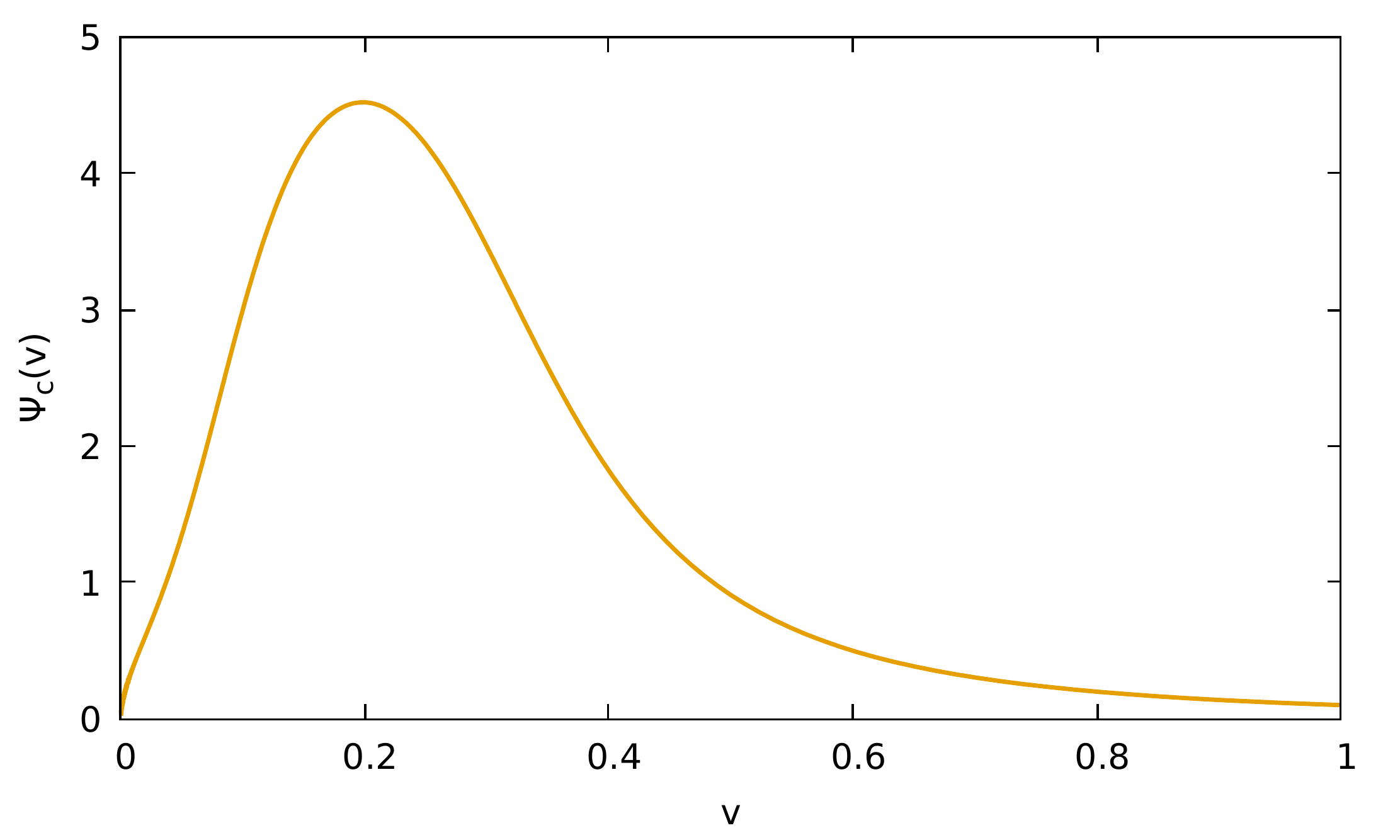}
   \end{center}
  \caption{Critical scaling function $\Psi_c ( \nu )$ of the dynamic structure factor
 $S ( \bd{k} , \omega )$ defined via Eq.~(\ref{eq:Psicdef}) for $d=3$.
}
\label{fig:Psicrit}
\end{figure}

Precisely at the critical point $ k \xi = \infty$ so that 
it is convenient to express the dissipation energy and the dynamic structure factor
 in terms of
$\nu = y / x^{z} = \omega / \omega_k$, see Eqs.~(\ref{eq:DeltaB}) and (\ref{eq:SPsi}). Setting $x = \infty$ in these expressions and defining the critical scaling functions 
\begin{eqnarray}
 B_c ( i \nu ) & = & B ( \infty , i \nu ),
 \\
 \Psi_c ( \nu ) & = & \Psi ( \infty , \nu ),
 \end{eqnarray}
the dissipation energy and the
dynamic structure factor at the critical point can be written as
  \begin{eqnarray}
 \Delta ( \bd{k} , i \omega ) & = &  \omega_k B_c (  i \omega / \omega_k ),
 \label{eq:Bcdef}
 \\
 S ( \bd{k} , \omega ) & = & \frac{ T_c G ( \bd{k} )}{\pi\omega_k} \Psi_c (  \omega / \omega_k ).
 \label{eq:Psicdef}
\end{eqnarray}
In Figs.~\ref{fig:Bcrit} and \ref{fig:Psicrit} we show our results for the critical 
scaling functions 
$B_c ( i \nu )$ and $\Psi_c ( \nu )$ in three dimensions.
Taking the limit $x \rightarrow \infty$ 
in the integral equation (\ref{eq:Bint1}) for the scaling function $B ( x , i \nu )$
we find that for $d = 3$
the critical scaling function $B_c ( i  \nu )$ satisfies the integral equation
 \begin{eqnarray}
 B_c ( i \nu ) & = & \frac{1}{ 2\pi^2 }  \int_0^{\infty} d \rho
 \frac{1}{ \rho^z B_c ( i \nu / \rho^z ) + | \nu | }
 \nonumber
 \\
 &  \times & \left[ 1 + \frac{\rho^3}{2} \ln \left| \frac{ \rho +1 }{\rho -1 }
 \right| - \rho^2 \right],
 \label{eq:Bcintegral}
 \end{eqnarray} 
where $z =5/2$. 
The asymptotic behavior of $B_c ( i \nu)$
for small and large $\nu$ can be  easily obtained
without explicitly solving the  
integral equation (\ref{eq:Bcintegral}).
Keeping in mind that the integral in Eq.~(\ref{eq:Bcintegral}) is cut for
$ \rho \lesssim | \nu |^{1/z}$, we obtain in the regime $| \nu | \ll 1$  to leading order
 \begin{eqnarray}
  B_c ( i \nu )  & \sim &  \frac{1}{ 2 \pi^2 }  \int_0^{\infty} d \rho
 \frac{1}{ \rho^z B_c ( i \nu / \rho^z ) + | \nu | }
 \nonumber
 \\
   & = & \frac{ | \nu |^{1/z -1} }{ 2 \pi^2 z} 
 \int_0^{\infty} ds \frac{s^{-1/z}}{ B_c ( i s \,  {\rm sgn} \nu  ) + | s | }
 \nonumber
 \\ 
 & = & 
 \frac{B_0} { | \nu |^{3/5}} , \; \; \; \mbox{for $| \nu | \rightarrow 0 $},
 \label{eq:Bcsmall}
 \end{eqnarray}
where
 \begin{equation}
 B_0 = \frac{1}{5\pi^2} 
 \int_0^{\infty} ds \frac{s^{-2/5 }}{ B_c ( i s \,  {\rm sgn} \nu  ) + | s | }.
 \end{equation}
In the opposite limit $ | \nu | \gg 1$ we may expand
the term in the last line of Eq.~(\ref{eq:Bcintegral}) to leading order for large $\rho$,
 \begin{equation}
1 + \frac{\rho^3}{2} \ln \left| \frac{ \rho +1 }{\rho -1 }
 \right| - \rho^2 = \frac{4}{3} + {\cal{O}} ( 1 / \rho^2 ) .
 \end{equation}
It follows that for large $| \nu |$ the scaling function $ B_c ( i \nu )$ has a similar
asymptotic behavior as for small $ | \nu |$ with a different prefactor,
 \begin{equation}
 B_c ( i \nu ) \sim \frac{4}{3}  
 \frac{B_0} { | \nu |^{3/5}} , \; \; \; \mbox{for $| \nu | \rightarrow \infty $}.
 \label{eq:Bclarge}
 \end{equation}
According to Eq.~(\ref{eq:PsiBscale})
the corresponding line-shape of the dynamic structure factor 
is given by the critical scaling function
  \begin{equation}
 \Psi_c ( \nu ) = \frac{1}{\nu} {\rm Im}
 \left[ \frac{ B_c (  \nu + i 0 )}{ B_c (  \nu + i 0) - i \nu } \right].
 \label{eq:PsiBscalecrit}
 \end{equation}

Our result~(\ref{eq:Bclarge}) for large $ | \nu |$ implies that for large frequencies
 $ \omega \gg \omega_k \propto k^{5/2}$
the dynamic structure factor at the critical point of a three-dimensional ferromagnet 
exhibits a non-Lorentzian decay,
 \begin{equation}
 S ( \bd{k} , \omega ) \propto \omega^{ -13/5}, \; \; \; \omega \rightarrow \infty,
 \label{eq:Stail}
 \end{equation}
in agreement with previous calculations \cite{Wegner68,Bhattacharjee81}.
Note that a Lorentzian line-shape decays for large frequencies as $\omega^{-2}$,
implying a larger tail than predicted by Eq.~(\ref{eq:Stail}).
The non-Lorentzian high-frequency tail can also be observed for $T > T_c$
in the regime $ ( \omega/\omega_\ast)^{1/z} 
 \gg \max \{ (\xi/a)^{-1}, ka \}$, which is equivalent with the condition 
$\nu^{2/5} \gg \max \{ x^{-1}, 1 \}$, as can be inferred 
from the integral equation~(\ref{eq:Bint1})  for $B ( x , i \nu )$.

On the other hand, for small frequencies our  result (\ref{eq:Bcsmall}) implies
 \begin{equation}
 S ( \bd{k} \neq 0 , \omega ) \propto \omega^{ 3/5}, \; \; \; \omega \rightarrow 0,
 \label{eq:Ssmall}
 \end{equation}
which contradicts previous findings from mode-coupling calculations \cite{Wegner68,Hubbard71,Frey94} and perturbative RG calculations~\cite{Bhattacharjee81,Folk85,Folk06} using an extrapolation of a truncated 
expansion in powers of $\epsilon = 6-d$ to the physically relevant case $\epsilon =3$.
Both methods predict a finite value of $S ( \bd{k} , 0 )$ at the critical point, corresponding to a finite limit of the scaling function $B_c ( i \nu )$ for $\nu \rightarrow 0$.
The  non-analytic behavior of $B_c ( i \nu )$ given in Eq.~(\ref{eq:Bcsmall})
prediced by our approach leads to a broad hump in the spectral line-shape
centered at $\nu \approx 0.2$  and a pseudogap \cite{Timusk99} for smaller frequencies, as shown
in Fig.~\ref{fig:Psicrit}. Note that the concept of a 'pseudogap' has been used in the literature on high-temperature superconductors to describe the low-frequency suppression of spectral weight observed in magnetic scattering \cite{Timusk99}. In the same sense the low-frequency feature described by \eqref{eq:Ssmall}, i.e. the (non-analytic) suppression of $S(\bm{k},\omega)$ for $\omega \rightarrow 0$, can be called a pseudogap.

\begin{figure}[tb]
 \begin{center}
  \centering
\vspace{7mm}
    \includegraphics[width=0.45\textwidth]{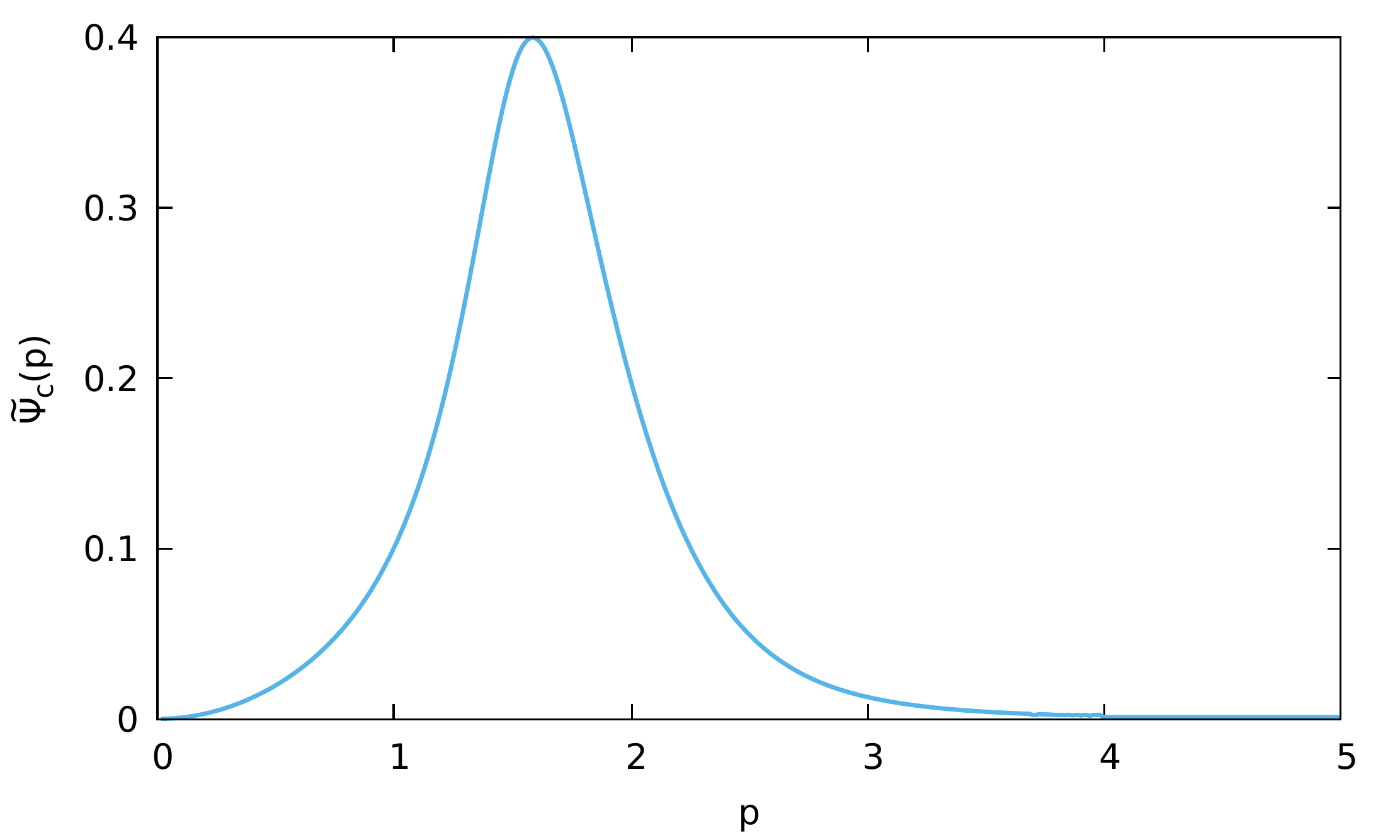}
   \end{center}
  \caption{Scaling function $\tilde{\Psi}_c ( p ) $ defined in Eq.~(\ref{eq:tildePsip}) describing 
the momentum dependence of the dynamic structure factor at
the critical temperature.}
\label{fig:crit_shape_mom_dep}
\end{figure}

The corresponding momentum dependence of $S(\bm{k},\omega)$ at the critical point 
is shown in Fig. \ref{fig:crit_shape_mom_dep}. It is convenient to measure momenta in units of $k_{\omega}$ defined by
 \begin{equation}
 \nu = \frac{\omega}{\omega_k} = \left( \frac{ k_{\omega}}{k} \right)^z,
 \end{equation}
which is equivalent with
 $k_{\omega} a =  ( \omega/\omega_\ast)^{1/z} $.
From Eq.~(\ref{eq:Psicdef})
we see that the momentum dependence of the dynamic structure factor at the
critical point is proportional to the scaling function
 \begin{equation}
 \tilde{\Psi}_c ( p ) = p^{-9/2} \Psi_c ( p^{-2/5} ),
 \label{eq:tildePsip}
 \end{equation}
where $ p = k / k_{\omega} = \nu^{-1/z}$.
Due  to the 
conservation of total spin, for fixed $\omega \neq 0$ 
the dynamic structure factor exhibits a peak at finite momentum. The asymptotic behavior for $\nu \ll 1 $ implies that 
\begin{equation}
 S ( \bd{k} , \omega \neq 0 ) \propto k^2, \; \; \; k \rightarrow 0,
 \label{eq:Sksmall}
 \end{equation}
in agreement with previous calculations 
\cite{Wegner68, Hubbard71, Bhattacharjee81,Folk85,Frey94,Folk06}.
Note that a Lorentzian profile would imply $S(\bm{k},\omega) \propto k^{1/2}$. 
On the other hand,
for large momenta  $k \gg k_\omega$ we find that $S ( \bd{k} , \omega )$ decays as
$k^{-6}$ which disagrees with the $k^{-9/2}$ behavior corresponding to a Lorentzian and
the more intricate line-shapes of mode-coupling theory and perturbative RG calculations \cite{Wegner68, Hubbard71,Bhattacharjee81,Folk85,Frey94,Folk06}.

\subsection{Comparison with previous calculations}

As already shown in Fig.~\ref{fig:Psicrit}, the pseudogap in the spectral line-shape
$ \omega \ll \omega_k$ with the non-analytic frequency dependence
given in  Eq.~(\ref{eq:Ssmall}) disagrees with the results of previous  theoretical investigations
using either mode-coupling theory \cite{Wegner68,Hubbard71,Frey94} or 
perturbative  RG methods \cite{Bhattacharjee81,Folk85,Frey94,Folk06},
based on a low-order expansion in powers of
$\epsilon = 6-d$. Both methods give a finite value of $S ( \bd{k} , 0 )$; in fact, from the extrapolation of a RG calculation using a  truncated $\epsilon$-expansion to order $\epsilon^2$  
it has been found 
that $S ( \bd{k} , \omega )$ assumes a unique maximum at vanishing 
frequency \cite{Bhattacharjee81}. 
A possible explanation of this discrepancy is that, at least for $d = 3$, our approach is simply
not valid for small frequencies $ \omega \ll \omega_k$, so that the pseudogap 
in the spectral line-shape predicted by our approach in this regime
is an unphysical artefact of our approximate method.
On the other hand, mode-coupling theory also uses  a number of  uncontrolled approximations, 
while the validity of the extrapolation of a low-order expansion in powers of $\epsilon = 6-d$
to the physically relevant case $\epsilon =3$  is questionable.

In principle, this problem can be clarified by means of large-scale numerical simulations.
Unfortunately, numerical spin dynamics calculations of 
the critical line-shape of isotropic ferromagnets performed many years ago by Chen 
and Landau \cite{Chen94} do not  cover the relevant regime of
arbitrarily small momenta. 
More recent numerical spin dynamics results are available only outside the scaling 
regime~\cite{Tao05} and give evidence for the existence of 
well-defined paramagnetic spin waves with sufficiently short wavelengths 
in Heisenberg ferromagnets. In this context a one-peak structure at long wavelengths, in line with previous investigations, was indeed mentioned by the authors of \cite{Tao05} although an actual line-shape was never shown.
Apparently the results for the scaling regime were at this point assumed to be converged and interest in this particular problem has waned. 
In a subsequent review \cite{Folk06} on critical dynamics no allusion to this calculation was made and the statement of \cite{Chen94} regarding the absence of a controlled numerical result for $S(\bm{k},\omega)$ was repeated again. 
In spite of the situation being still somewhat ambiguous, 
it is fair to say that so far there is no numerical evidence supporting the non-analytic vanishing of $S(\bm{k},\omega)$ for $\omega \rightarrow 0$ obtained by our calculation.
This suggests that the pseudogap feature is simply an artifact of our method.
Nevertheless, as shown in the following subsection, in spite of the (perhaps) unphysical
pseudogap feature, our approach leads to a satisfactory agreement with available experiments.

\subsection{Comparison with experiments}
\label{sec:experiments}

In Refs.~[\onlinecite{Boeni86,Boeni87}] experimental  results
for the neutron scattering cross-section at the critical temperature 
$T_c = 69.25 \ \mathrm{K}$ 
of the magnetic insulator EuO have been presented. This material is well described by a Heisenberg ferromagnet with nearest and next-nearest neighbor exchange interactions 
$J_1 = 1.21 \, \mathrm{K}$ and $ J_2 =  0.24 \, \mathrm{K}$ \cite{Passell76} on a face-centered cubic (fcc) lattice with lattice spacing $a = 5.12 \ \text{\AA}$. Note that for a fcc lattice the ratio of the volume of the primitive unit cell  to the volume of the conventional unit cell is $v = 1/4$.
To compare the  experimental data with our theoretical predictions presented above, we should take the finite energy resolution $\delta_{\omega}$ of the experiment into account.
This can be achieved by convoluting our theoretical prediction for
$S ( \bd{k} , \omega )$ with the experimentally relevant 
resolution function $E ( \omega )$ such that the experimentally measured neutron 
scattering cross section is proportional to
\begin{equation}
 S_{\rm con } (\bm{k},\omega) = \int_{-\infty}^{\infty}d\omega' E(\omega - \omega')S(\bm{k},\omega').
\label{eq:convol_dynstruc}
\end{equation}
Usually $E(\omega )$ is chosen to be a Gaussian with width $\delta_{\omega}$, 
\begin{equation}
E(\omega) = \frac{1}{\sqrt{2\pi\delta^2_\omega}}\exp\Big[ - \frac{\omega^2}{2\delta^2_\omega}\Big]. 
\label{eq:energy_res_func}
\end{equation}
The experimental resolution in the experiment  by B\"{o}ni {\it{et al.}} \cite{Boeni87} is 
$\delta_\omega = 0.05 \ \mathrm{meV}$. Intuitively it is clear that the pseudogap for
$\omega \lesssim \omega_k$ predicted by our theory can only be resolved experimentally if
the relevant energy scale $\omega_k$ is large compared with the experimental resolution
$\delta_{\omega}$. Below we show that this is not the case, so that the experimental data
of Ref.~[\onlinecite{Boeni87}] cannot resolve a possible pseudogap.

\begin{figure}[tb]
 \begin{center}
  \centering
\vspace{7mm}
   \includegraphics[width=0.45\textwidth]{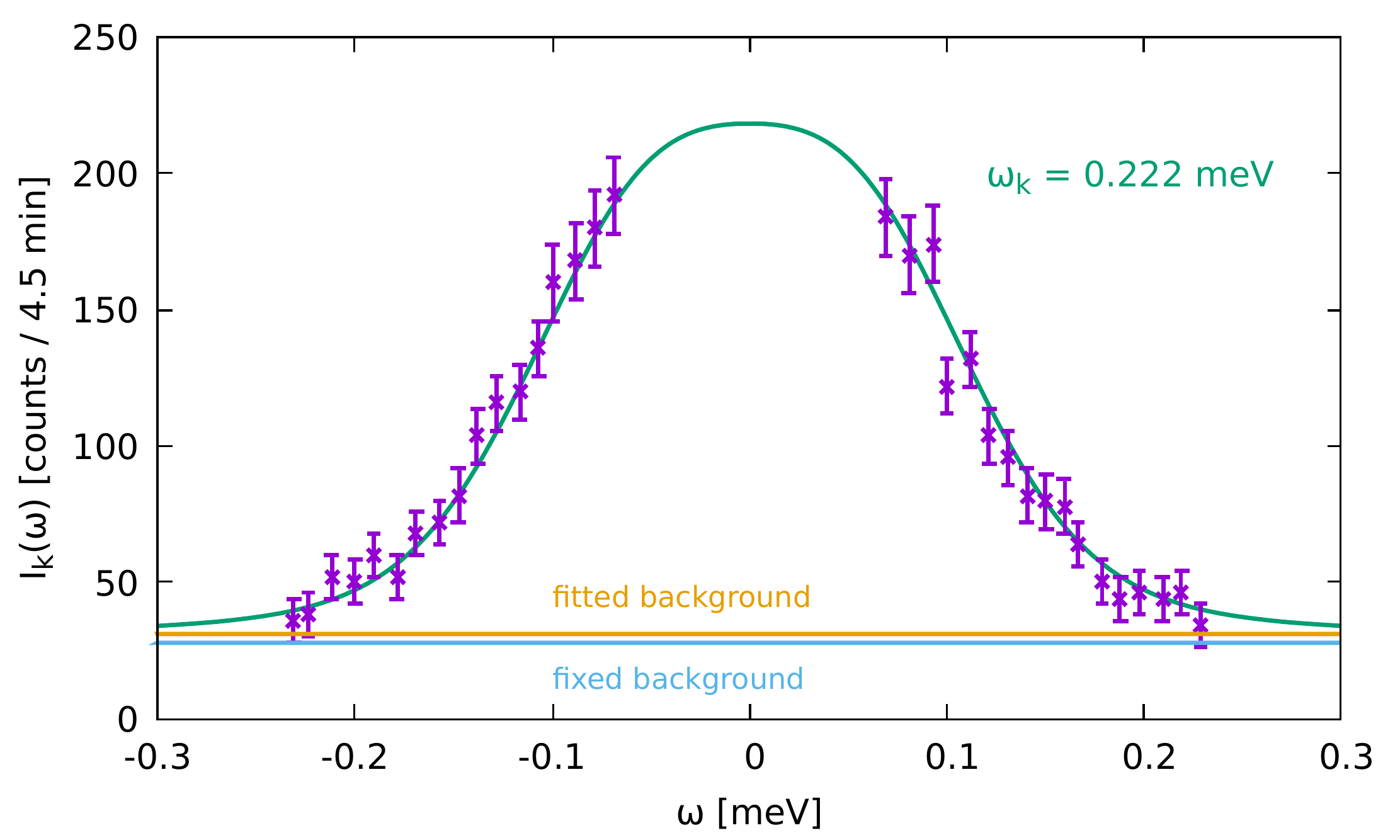}
   \end{center}
  \caption{Fit of our theoretical prediction
for the convoluted neutron scattering intensity given by  Eq.~\eqref{eq:fit_mom_const} to the experimental neutron scattering intensity at  constant wavevector  
$k = 0.15 \ \text{\AA}^{-1}$ displayed in Fig.~2 of  Ref. \cite{Boeni87}.
As discussed in the text,  we have omitted data at 
small frequencies $\omega \lesssim \Gamma_k = 0.072 \ \mathrm{meV}$. 
Since in Ref.~[\onlinecite{Boeni87}] the experimental error was  given only for a few points,
we have estimated the remaining errors by assigning 
identical error bars  to groups of adjacent points.
}
\label{fig:fit_const_mom_without_small_freq}
\end{figure}

Following the procedure described by B\"{o}ni {\it{et al.}} \cite{Boeni87} where elastic ($\omega = 0$) nonmagnetic scattering is subtracted, 
we make the following ansatz for the experimentally observed neutron scattering intensity  at {\it{constant momentum}},
\begin{equation}
\mathcal{I}_k (\omega) = C  S_{\rm con} (\bm{k},\omega) + B.
\label{eq:fit_mom_const}
\end{equation}
Here the normalization constant $C$ and  the background $B$ are fit parameters;
moreover, we use also the characteristic energy scale $\omega_k$ in our scaling functions
as a fit parameter.
After fixing $C$ and $\omega_k$ via a $\chi^2$- fit we compare the data 
{\it{at constant frequency}} $\omega$  with
\begin{equation}
\mathcal{I}_\omega(k) = C S_{\rm con} (\bm{k},\omega) + B',
\label{eq:fit_freq_const}
\end{equation}
where the backgound $B'$ is the only remaining  free parameter. 
In Fig.~\ref{fig:fit_const_mom_without_small_freq}  we show a fit 
of our result for the convoluted neutron scattering intensity given by
 Eq.~\eqref{eq:fit_mom_const} to the measured data displayed in Fig.~2 of Ref.~\cite{Boeni87} 
with $k = 0.15 \ \text{\AA}^{-1}$. 
Given our expectations regarding the applicability of our results, for the fit we have included only
 data with frequencies in the regime $\omega \gtrsim \Gamma_k$,
 where $\Gamma_k \approx 0.072 \ \mathrm{meV}$ is the experimentally determined line-width for a Lorentzian \cite{Boeni86,Boeni87,Resibois70}
\begin{equation}
S(\bm{k},\omega) \propto \frac{\Gamma_k}{\Gamma^2_k + \omega^2}.
\end{equation}
Note that in the analysis of the experimental data a heuristic modification of the simple Lorentzian was initially used \cite{Wicksted84, Boeni86} which separates the data by the same criterion in order to account better for the data at large frequencies. 
Later this ansatz was replaced by the interpolation formula from asymptotic renormalization group calculations \cite{Bhattacharjee81,Folk85}, which in fact predicts a shape similar to the empirical ansatz.
Obviously, the convoluted spectral line-shape in Fig.~\ref{fig:fit_const_mom_without_small_freq} exhibits only a central peak, i.e. the
pseudogap for small frequencies predicted by our theory is not visible due to the rather
large experimental resolution. The obtained value for the characteristic frequency $\omega_k = 0.222 \ \mathrm{meV}$ is somewhat larger than our theoretical prediction $\omega_k = 0.158 \ \mathrm{meV}$ if we use 
accepted values for the bare stiffness 
$J'' = J_1 + J_2 = 1.45 \ \mathrm{K}$.
The background $B \approx 31 \ \mathrm{counts}$  is not too far off from the fixed value $B = 28$, where the latter is extracted from the measured scattering  at low temperatures \cite{Boeni87} and is also obtained by fitting the data to the result of  asymptotic renormalization group theory  \cite{Bhattacharjee81,Folk85}. In fact, the fit parameters $C$ and $ \omega_k$ do not significantly change if we choose $B =28$. We conclude that the experimental line-shape for not too small frequencies is 
quantitatively explained by our theory. In contrast, an attempt to fit the data with a Lorentzian line-shape overestimates the spectral weight in the high-frequency tails.

\begin{figure}[tb]
 \begin{center}
  \centering
\vspace{7mm}
    \includegraphics[width=0.45\textwidth]{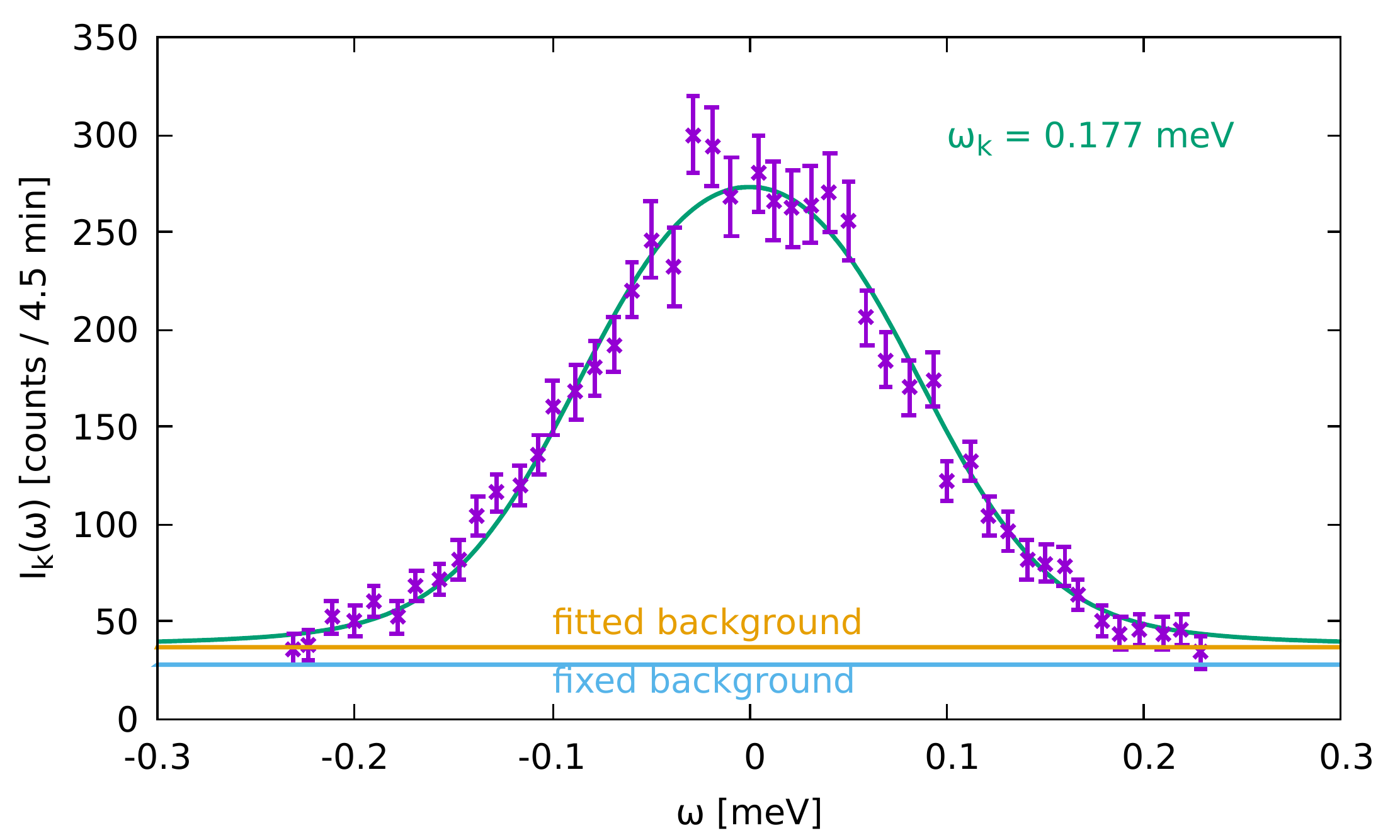}
   \end{center}
  \caption{Fit of our theoretical prediction for the convoluted neutron scattering intensity 
given by Eq. \eqref{eq:fit_mom_const} to the data 
at constant wavevector $k = 0.15 \ \text{\AA}^{-1}$ displayed in Fig.~2 of Ref. \cite{Boeni87}.}
\label{fig:fit_const_mom}
\end{figure}

For completeness we also show in Fig.~\ref{fig:fit_const_mom} a fit of our
theory to the full set of data.  
The characteristic frequency now comes out smaller $\omega_k = 0.177 \ \mathrm{meV}$ and
agrees much better with the theoretical value  $\omega_k = 0.158 \ \mathrm{meV}$.
The  larger ratio $\delta_\omega/\omega_k$ implies a smoother and slimmer 
lineshape around $\omega = 0$. 
The larger background $B \approx 37 \ \mathrm{counts}$ necessary for the fit
might indicate that for small frequencies our theoretical approach does not correctly account 
for the data. 

\begin{figure}[tb]
 \begin{center}
  \centering
\vspace{7mm}
  \includegraphics[width=0.45\textwidth]{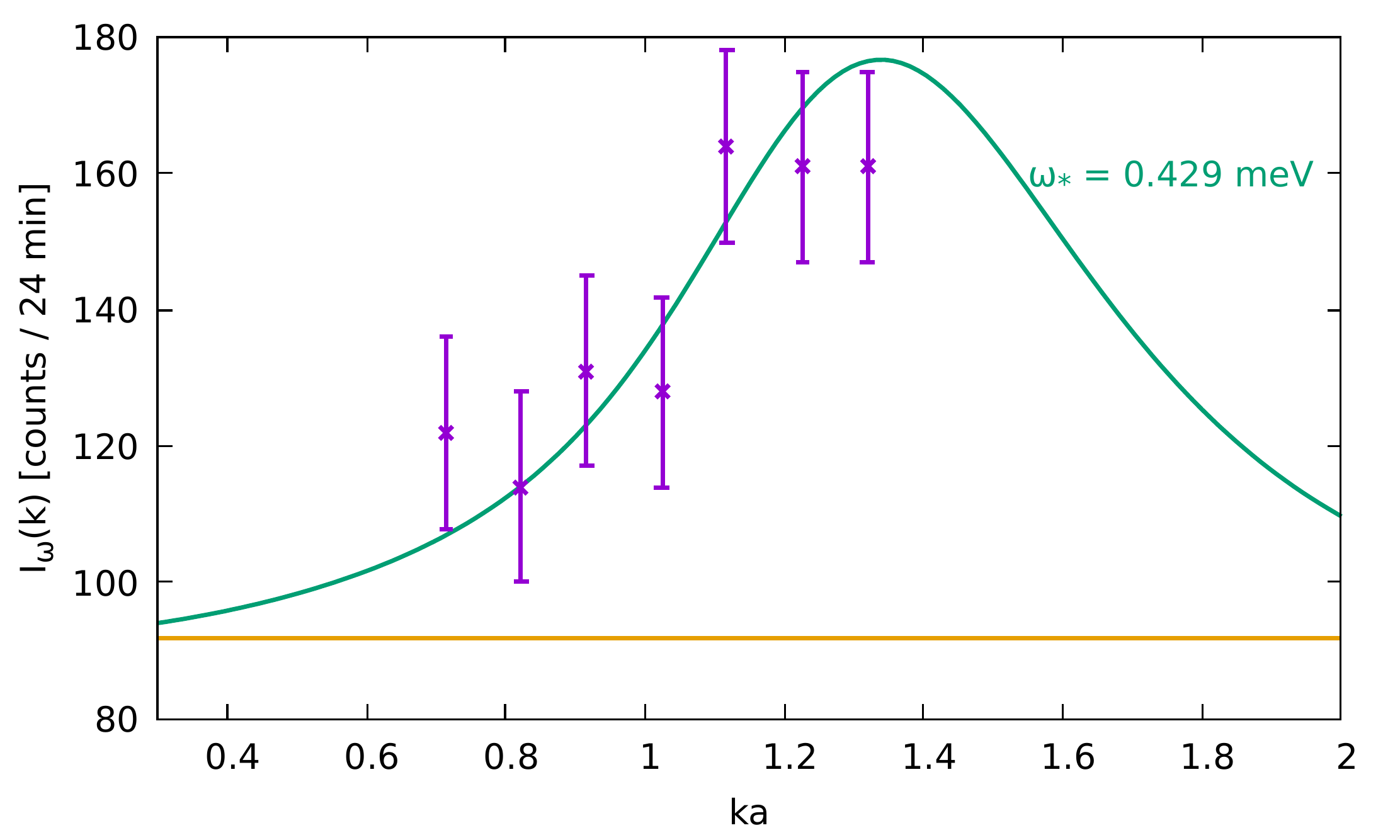}
   \end{center}
  \caption{Fit of our theoretical prediction for the convoluted intensity
given in  Eq. \eqref{eq:fit_freq_const} to experimental data at constant frequency $\omega = 0.3 \ \mathrm{meV}$ displayed in Fig.~3 of Ref. \cite{Boeni87}.  For consistency we have considered only 
data with small momenta 
$ka < (\omega/\Gamma')^{2/5} \approx 1.36$, because for large momenta $k \gtrsim k_\omega$ we expect a breakdown of our ansatz.
Note that the data shown here are collected during a larger time-interval than the data
in Fig.~\ref{fig:fit_const_mom}.
}
\label{fig:fit_const_freq_without_large_mom}
\end{figure}

Another way to test the consistency of our results with experiments is to compare
scans at constant energy transfer, described by Eq.~\eqref{eq:fit_freq_const}, which are much more sensitive 
to the precise form of the line-shape 
due to the shape-dependence of the peak position $k_\ast (\omega)$. 
In our case we find $k_\ast (\omega)a \approx [3.25 \, \omega/\omega_{\ast}]^{2/5} < 
k_\omega a$, where the microscopic energy $\omega_\ast$ is defined in Eq. \eqref{eq:omega_ast}, and a half-width $\Delta k(\omega) \approx 0.5 \, k_\ast(\omega)$. 
For a Lorentzian with $\Gamma' = \Gamma_k (ka)^{-5/2}$ (where $\Gamma' = 
0.139 \ \mathrm{meV}$ for EuO) 
the peak position is $k_\ast (\omega)a = [\omega/(3\Gamma')]^{2/5}$ 
with $\Delta k (\omega)/k_\ast(\omega) = 1.57$. 
This should be compared with the results  of asymptotic RG
calculations \cite{Folk85}, which give  for the peak position
$k_\ast (\omega)a = [\omega/(1.3\Gamma')]^{2/5} $ and for the width 
$\Delta k(\omega)/k_\ast(\omega)= 0.75$. 
Using our result from the fit to the scan at constant momentum 
we have $\omega_{\ast} = 0.429 \ \mathrm{meV} \approx 3 \Gamma'$, which implies  that our predicted peak positions are about 15~$\%$ larger than the RG result \cite{Folk85,Boeni87}, whereas for a Lorentzian the peak positions are about 30~$\%$ smaller.
In Fig.~\ref{fig:fit_const_freq_without_large_mom}  we show a fit of our theoretical prediction
to the data at fixed frequency $\omega = 0.3 \ \mathrm{meV}$ presented in
 Fig.~3 of Ref.~\cite{Boeni87},  where we restricted ourselves 
to small momenta $ka \lesssim (\omega/\Gamma')^{2/5}$ (corresponding to the upper limit
$ka  < 1.36$). 
The fitted background $B' \approx 92 \ \mathrm{counts}$ is reasonably 
close to the corresponding result $B' \approx 83 \ \mathrm{counts}$ reported 
in Fig.~3 of Ref.~\cite{Boeni87}. 
The data shown in Fig.~\ref{fig:fit_const_freq_without_large_mom} seem to be compatible
with our theory, in particular sufficiently far away from the peak. 
However the limited number of points and the relatively large statistical errors do not allow for a strong statement concerning the validity of our theory.
In fact, the measured intensity data at the smaller frequency $\omega = 0.2 \ \mathrm{meV}$, which is  also shown in Fig.~3 of Ref.~\cite{Boeni87}, with the same restrictions to the data, show significantly less agreement to our theory. 

\begin{figure}[tb]
 \begin{center}
  \centering
\vspace{7mm}
     \includegraphics[width=0.45\textwidth]{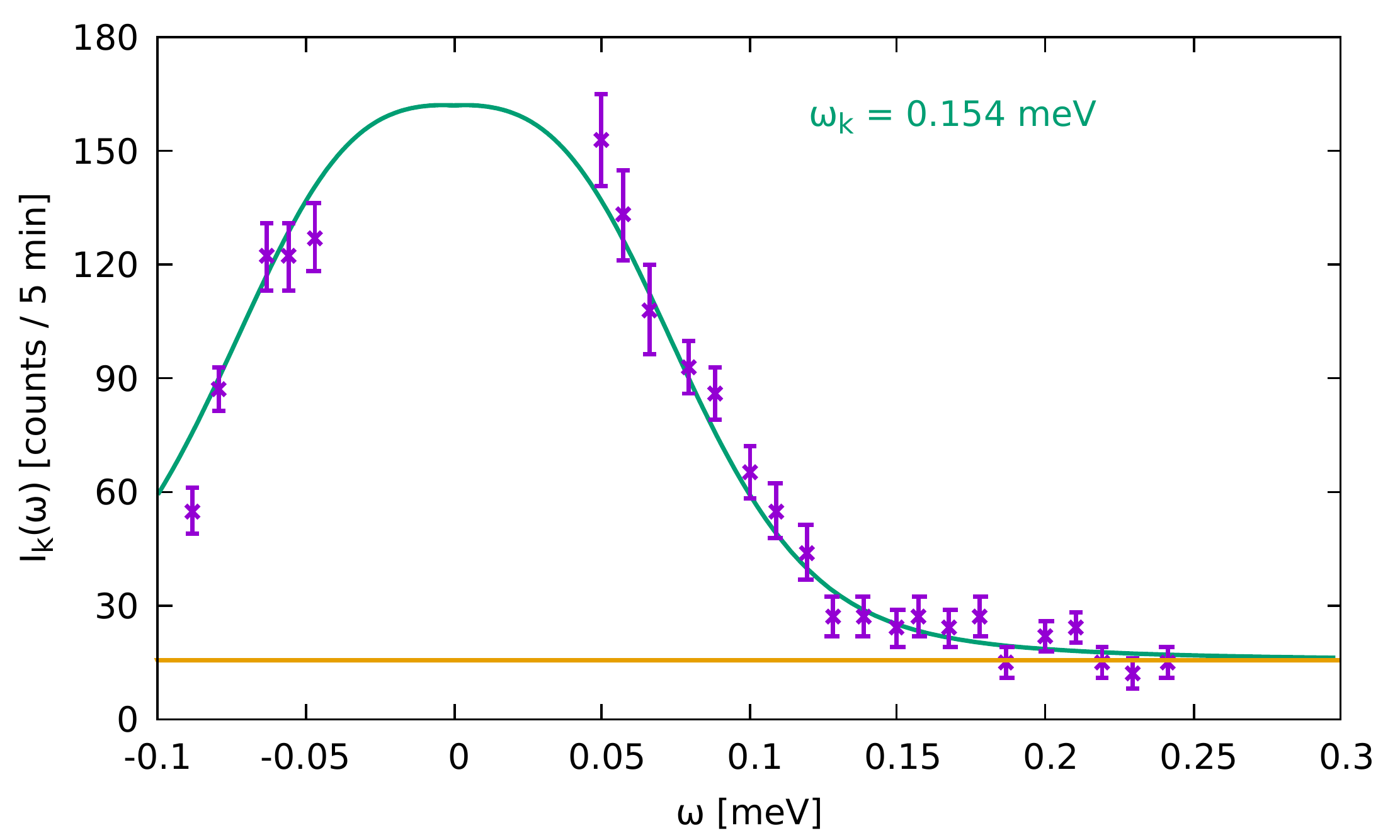}
   \end{center}
  \caption{Fit of our theoretical result \eqref{eq:fit_mom_const} for the convoluted line-shape to constant wavevector scan of the experimentally determined critical line-shape
of a EuS displayed in Fig.~2 of Ref.~[\onlinecite{Boeni88}]
at $k = 0.22 \ \text{\AA}^{-1}$. For the fit we have omitted
small frequencies $\omega \lesssim \Gamma_k = 0.051 \ \mathrm{meV}$ from the data.}
\label{fig:fit_const_mom_without_small_freq_II}
\end{figure}

\begin{figure}[tb]
 \begin{center}
  \centering
\vspace{7mm}
   \includegraphics[width=0.45\textwidth]{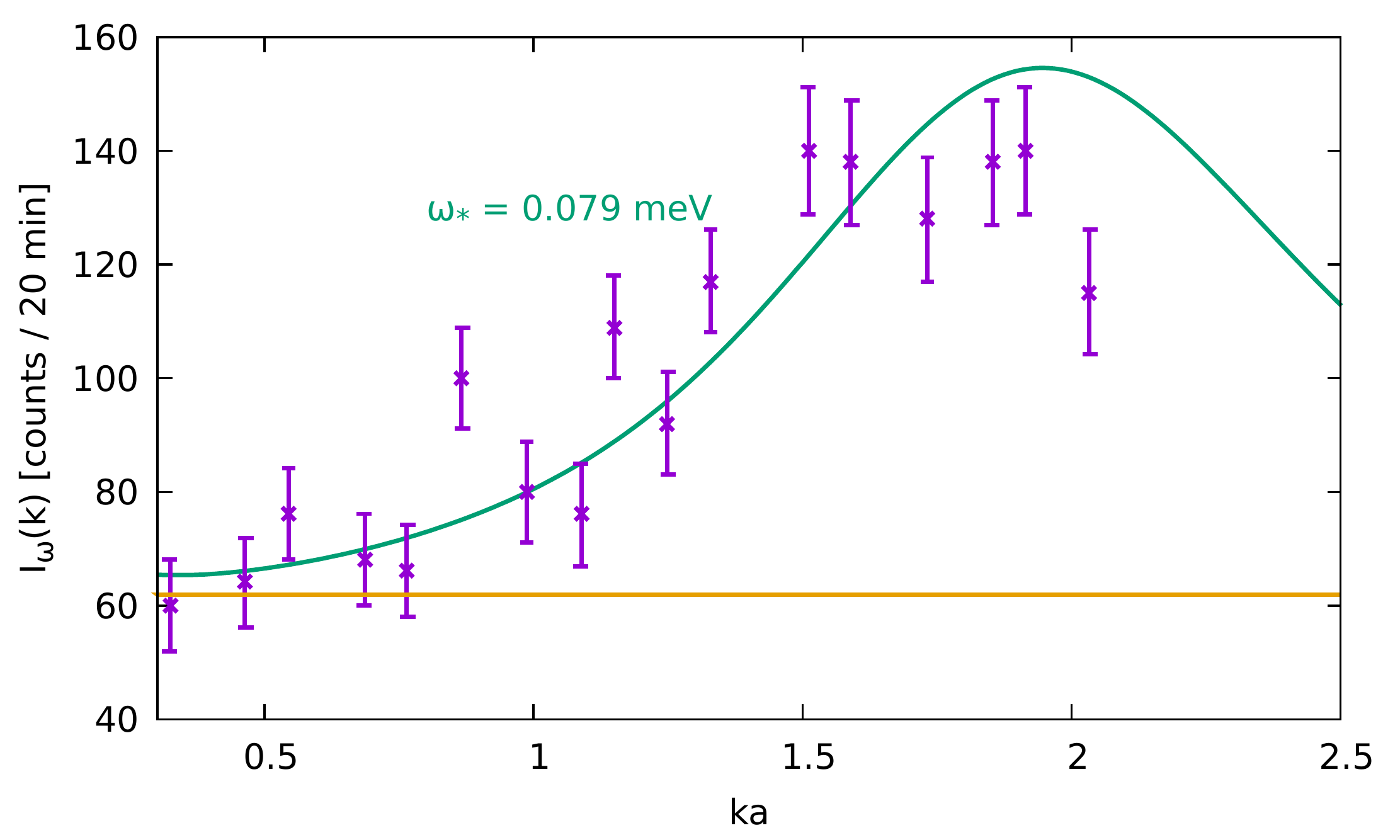}
   \end{center}
  \caption{Fit of our theoretical result  \eqref{eq:fit_freq_const} 
for the convoluted line-shape to constant frequency scan at $\omega = 0.15 \ \mathrm{meV}$ 
of the experimentally determined critical line-shape of EuS
displayed in Fig.~3 of Ref. \cite{Boeni88}.  For the fit we have 
dropped large momenta $ka \gtrsim (\omega/\Gamma')^{2/5} \approx 2.0 $.}
\label{fig:fit_const_freq_without_large_mom_II}
\end{figure}

Another material where high-precision neutron scattering data probing the critical spin dynamics are available is the related compound EuS \cite{Boeni88}, which is also a Heisenberg ferromagnet on a fcc lattice with lattice spacing $a = 5.95 \ \text{\AA}$, 
nearest-neighbor exchange  exchange interactions $J_1 = 0.47 \ \mathrm{K}$, next-nearest-neighbor exchange $  J_2 =  -0.24 \ \mathrm{K}$ \cite{Passell76}, 
and a critical temperature $T_c = 16.5 \ \mathrm{K}$. 
In Figs.~\ref{fig:fit_const_mom_without_small_freq_II} and \ref{fig:fit_const_freq_without_large_mom_II} we show fits of our theoretical results
to scans at the critical temperature reproduced from Fig.~2 and Fig.~3 
of Ref.~[\onlinecite{Boeni88}]. 
The experimental resolution is in this case
$\delta_\omega = 0.035 \ \mathrm{meV}$. The scan at fixed momentum shown
in Fig.~\ref{fig:fit_const_mom_without_small_freq_II} is for
$k = 0.22 \ \text{\AA}^{-1} $, while the constant-frequency scan in 
Fig.~\ref{fig:fit_const_freq_without_large_mom_II} is for
 $\omega = 0.15 \ \mathrm{meV}$. As in the case of EuO discussed above, 
we have retained only data  points which fulfill the respective conditions 
$\omega \gtrsim \Gamma_k = \Gamma'(ka)^{5/2} = 0.051 \ \mathrm{meV}$ or $ka \leq (\omega/\Gamma')^{2/5} \approx 2.0$, where $\Gamma' = 0.026 \ \mathrm{meV}$. 
From the  fixed momentum scan we obtain $\omega_\ast = 0.079 \ \mathrm{meV}$ 
which (like in EuO) is somewhat larger than the theoretically predicted value
 $\omega_\ast =  0.059 \ \mathrm{meV}$.
The fit itself agrees with the data quite well and
the fitted background $B$ coincides exactly  with the experimental value
$B = 15 \ \mathrm{counts}$.  Note that the ratio  between the fitted frequencies $\omega_\ast(\mathrm{EuO})/\omega_\ast(\mathrm{EuS})$ of both materials is consistent with $\Gamma'(\mathrm{EuO})/\Gamma'(\mathrm{EuS})$.
To our advantage, for the scan at fixed energy shown in 
Fig.~\ref{fig:fit_const_freq_without_large_mom_II}
much more points than for EuO \cite{Boeni87}
are now available to the left of the peak. The agreement between our theoretical prediction and the experimental data  is rather good, 
especially for the smaller momenta in this set. Stronger deviations start to appear only 
for the rightmost points, since the predicted peak position is again shifted by about 15 percent compared to the experimental value.

All in all, we have to admit that the interpolation formula from  a perturbative renormalization group calculation based on an extrapolation of a truncated $\epsilon = 6-d$-expansion \cite{Bhattacharjee81,Folk85},  
which was initially used to fit the data for a three-dimensional ferromagnet at $T=T_c$
by B\"{o}ni {\it{et al.}} \cite{Boeni87}, agrees with the 
experimental data somewhat better than the results obtained 
within our approach. This
might  indicate that the 
pseudogap of the critical line-shape for $\omega \lesssim \omega_k$ is an artefact of our approach, although  it is not visible in the spectral line-shape 
for small frequencies 
due to the finite experimental resolution.
On the other hand, for sufficiently large frequencies our approach produces critical line-shapes
which are fully consistent with the experimental data and certainly perform better
than a simple Lorentzian in the lower intensity part of the scattering cross-section. Note that the momenta for the scans at fixed $k$ are actually quite large, i.e. $ka = 0.768$ for EuO 
and $ka = 1.309$ for EuS; especially for EuS these values exceed the expected
boundaries of  the critical region where dynamic scaling should hold. 
Surprisingly, for the systems under consideration dynamic scaling seems to be valid in a much larger range of momenta and frequencies \cite{Boeni87, Boeni88} than theoretically expected.

Let us conclude this section with a caveat.
In real magnets the spins are not only coupled by short-range exchange interactions,
but also be long-range dipole-dipole interactions, which we have completely ignored in our work because typically these are much smaller than the exchange interactions.
Nevertheless, for temperatures close to $T_c$ dipole-dipole interactions
cannot be neglected because they violate spin-conservation,
and are expected to  change the universality class. 
As a consequence, the dynamic exponent is expected 
to cross over from $z = 5/2$ to $z=2$ so that
also the linewidth $\Gamma_k$ and the characteristic frequency  $\omega_k$ should change for very small momenta. This is accompanied by a crossover in the spectral
line-shape, which was experimentally found in EuO at $T = T_c$ 
to be parametrized by a Lorentzian \cite{Mezei86} instead of the exchange-only result
for very small wavevectors $k$ (which are much smaller than the wavevectors 
probed in the experiments by B\"{o}ni {\it{et al.}} \cite{Boeni86,Boeni87}), although the change
in the linewidth still could not be discerned.
In fact, within mode-coupling theory the line-shape including 
dipolar interactions has been calculated by Frey {\it{et al.}}~\cite{Frey94, Frey89}
who found that the 
crossover scale for the linewidth is smaller by almost an order of magnitude compared 
to the crossover scale for the line-shape. One therefore has to be careful in applying results for pure Heisenberg systems to the small momentum-tail in constant energy scans.

\section{Low-temperature behavior of the dynamic structure factor
in reduced dimensions}
\label{sec:reduced}

For the calculations of the dynamic structure factor in the critical regime of  three-dimensional Heisenberg magnets  presented in Sec.~\ref{sec:3d} we have neglected the momentum dependence of the static self-energy $\Sigma ( \bd{k} )$ in the kernel $V ( \bd{k} , \bd{q} )$ of our integral 
equation (\ref{eq:Deltaint}) for the dissipation energy $\Delta ( \bd{k} , i \omega )$, see
Eq.~(\ref{eq:sigmaapprox}). We have justified this approximation by arguing that
the momentum dependence of the self-energy can be neglected due to the small value
of the anomalous dimension $\eta \approx 0.027$ of the Heisenberg model in $d=3$.
Obviously, in one and two dimensions this argument is not valid because
in this case the Heisenberg model does not have a critical point at finite temperature.
The momentum dependence of the static self-energy is then important and cannot be neglected. In fact, in order to obtain results which are compatible with dynamic scaling 
the kernel $V ( \bd{k} , \bd{q} )$ of our integral equation (\ref{eq:Deltaint}) has to be modified
by replacing the difference of the bare couplings  in the vertex-correction factor $Z ( \bd{q} , \bd{k} )$
defined in Eq.~(\ref{eq:Zdef}) 
by the difference of inverse static propagators,
 \begin{equation}
 J ( \bd{q} + \bd{k} ) - J ( \bd{q} )  \rightarrow G^{-1} ( \bd{q} + \bd{k} ) - G^{-1} ( \bd{q} ).
 \label{eq:JG}
 \end{equation}
A similar substitution has been proposed by 
Frey and Schwabl \cite{Frey94}
within mode-coupling theory to obtain the correct dynamic exponent $z = (5 - \eta)/2$  for $\eta \neq 0$ in the dynamic scaling law. Note that 
within an approximation where the 
momentum-dependence of the self-energy is neglected, the substitution (\ref{eq:JG}) has no effect.
The kernel $V ( \bd{k} , \bd{q} )$ in   our integral equation (\ref{eq:Deltaint}) for the dissipation energy is then replaced by
 \begin{eqnarray}
 {V}_\ast ( \bd{k} , \bd{q} ) & = &  \frac{T}{4}   G^{-1} ( \bd{k} ) G ( \bd{q} )  
 G ( \bd{q} + \bd{k} ) 
 \nonumber
 \\
  & \times &  [ G^{-1} ( \bd{q} ) - G^{-1} ( \bd{q} + \bd{k} )  ]^2  + ( \bd{k} \rightarrow - \bd{k} ) . 
 \hspace{7mm}
 \label{eq:Vmod}
 \end{eqnarray}
 
In dimensions $d \leq 2$ the Heisenberg model does not have any 
long-range magnetic order at finite temperature so that $T_c=0$ and the correlation length $\xi$ diverges for $T \rightarrow 0$.
To solve our integral equation (\ref{eq:Deltaint}) 
for the dissipation energy we need the 
static spin-spin correlation function $G ( \bd{k} )$  as an input, the calculation of which in reduced dimensions is by itself a challenging theoretical problem.

\subsection{One dimension}
 
\begin{figure}[tb]
\begin{center}
 \centering
\vspace{7mm}
\includegraphics[width=0.45\textwidth]{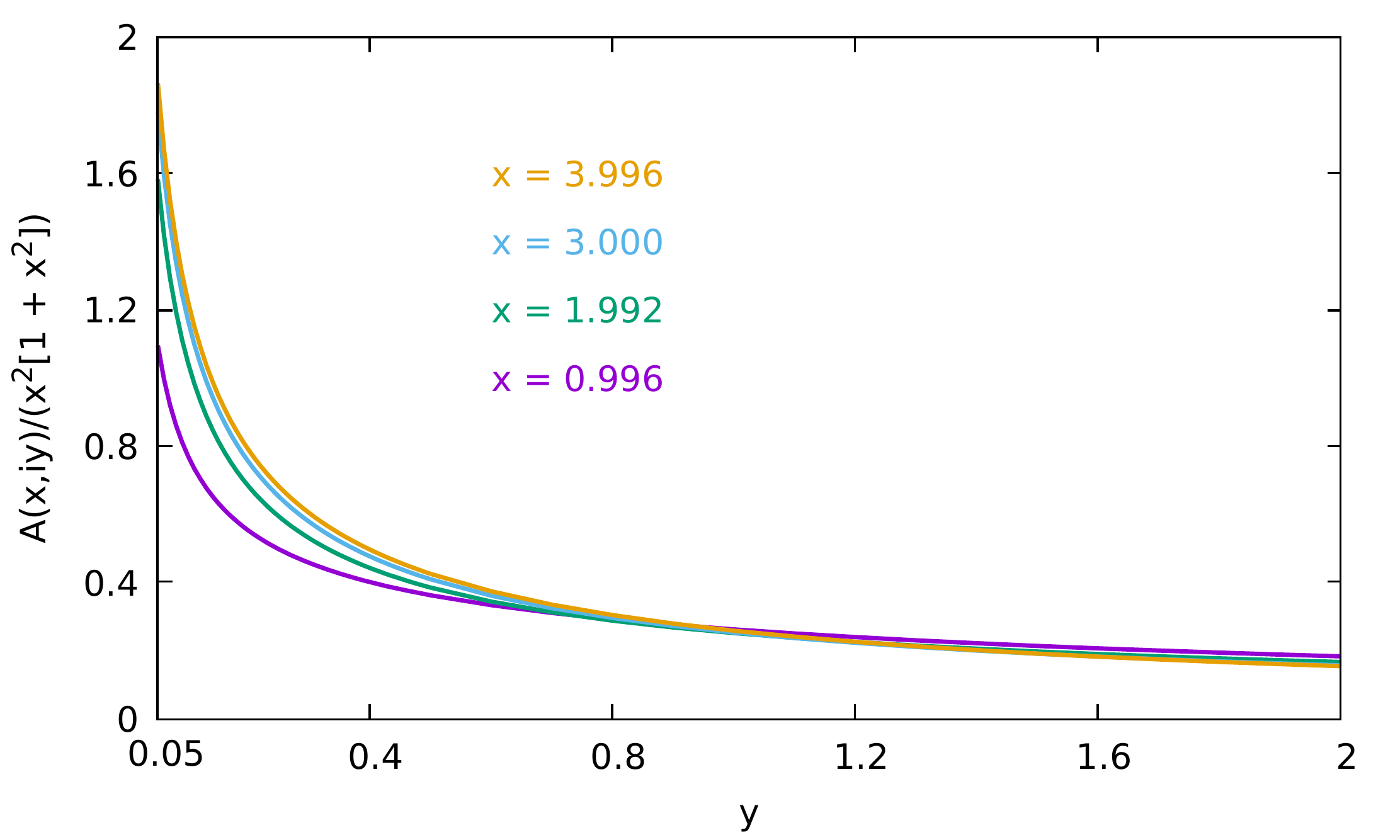} 
  \end{center}
  \caption{Frequency dependence of the scaling function $ A (x,iy)$ 
of the dissipation energy $\Delta (\bm{k} , i \omega )$
in one dimension obtained from the numerical solution of the
integral equation (\ref{eq:integralA1}). As in  Fig.~\ref{fig:SAres} we plot $A(x,iy)/(x^2[1+x^2])$ because this ratio depends only weakly on $x$.
}
\label{fig:Aplot1}
\end{figure}

\begin{figure}[tb]
\begin{center}
 \centering
\vspace{7mm}
\includegraphics[width=0.45\textwidth]{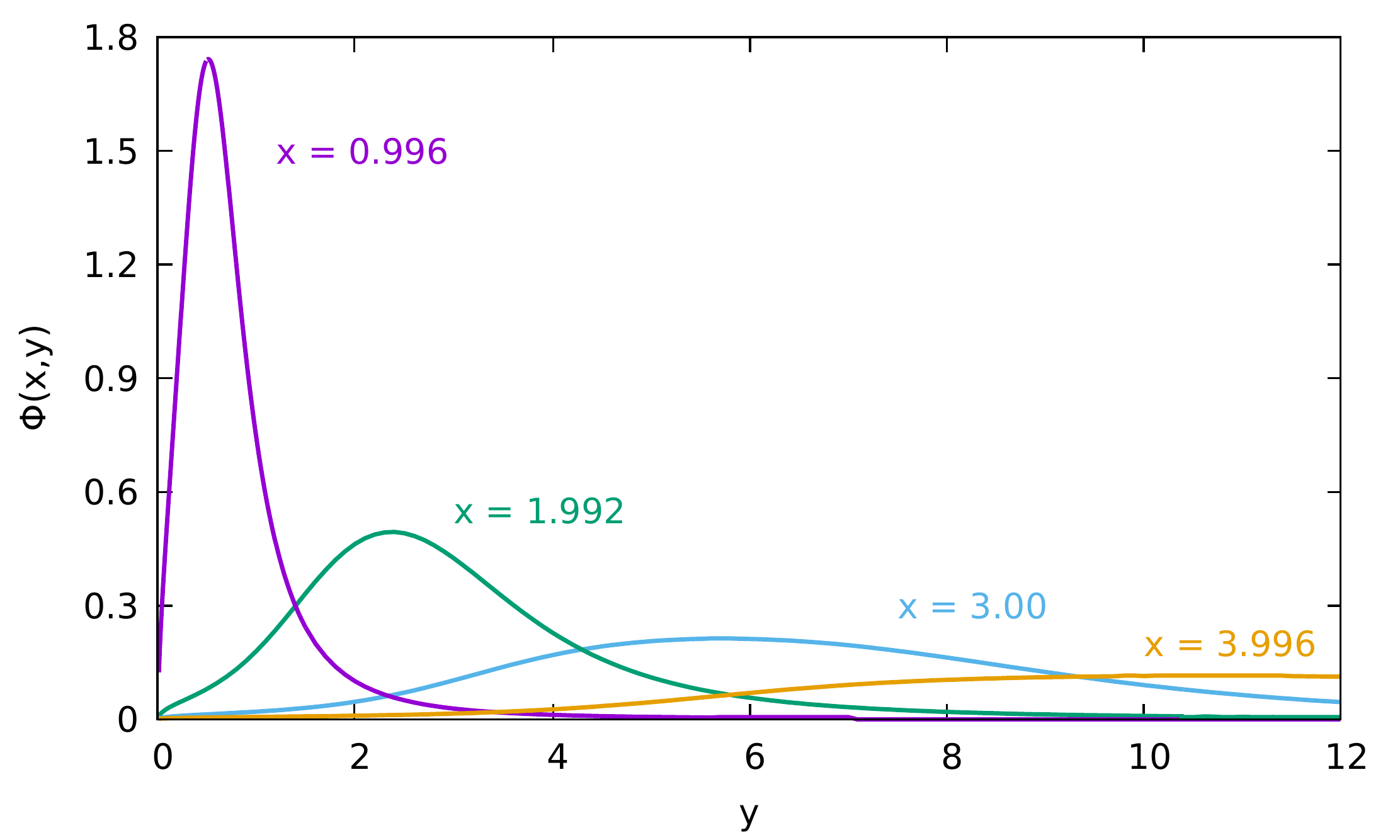}
   \end{center}
  \caption{Frequency dependence of the scaling function $\Phi(x,y)$ of the dynamic structure factor for a ferromagnetic Heisenberg chain
defined via Eqs.~(\ref{eq:Sscale1}) and (\ref{eq:phiscale})
for different values of $x = k \xi$.}
\label{fig:Phiplot1}
\end{figure}

\begin{figure}[tb]
\begin{center}
 \centering
\vspace{7mm}
\includegraphics[width=0.45\textwidth]{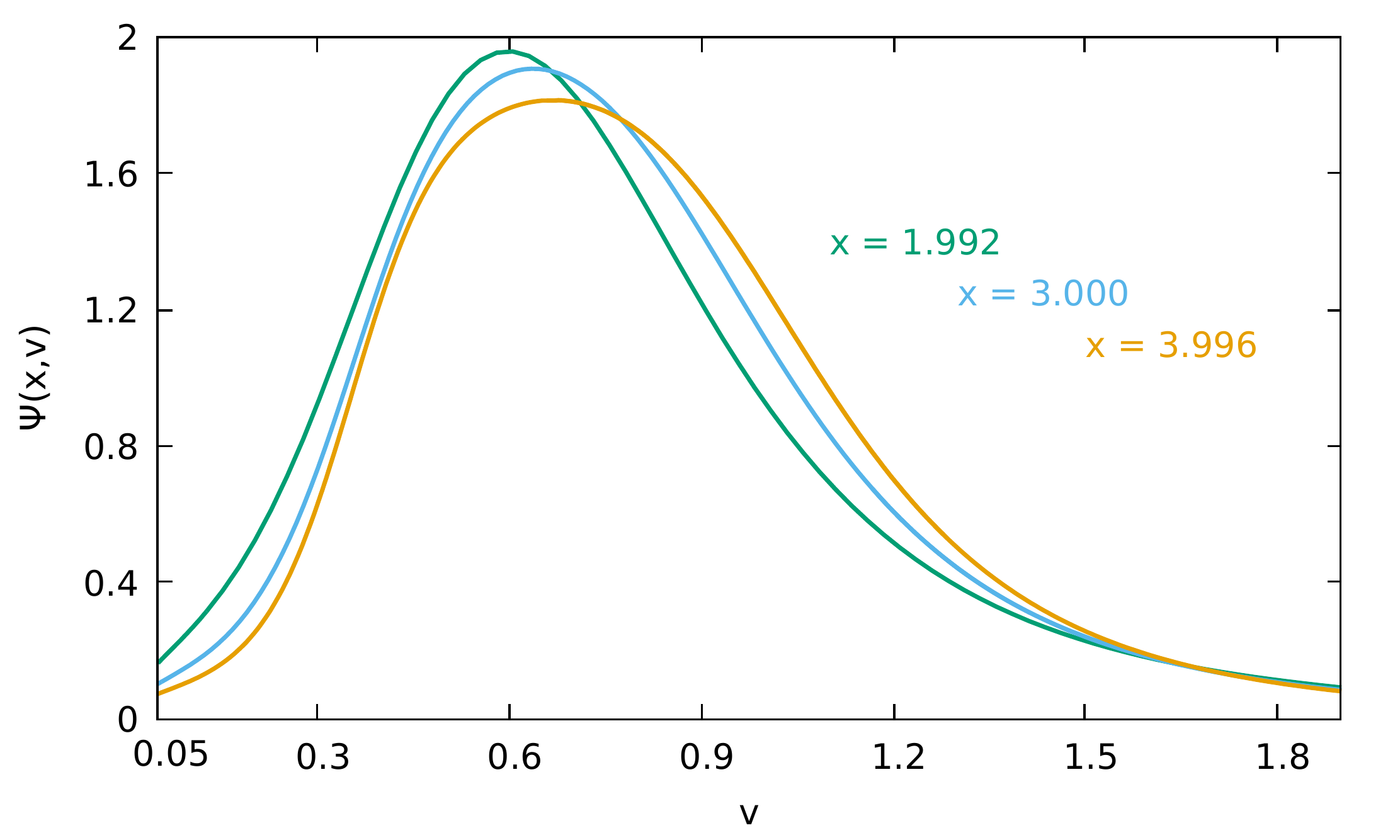}
   \end{center}
  \caption{Frequency dependence of the scaling function $ \Psi(x, \nu)$ of the dynamic structure factor for a ferromagnetic Heisenberg chain
for different values of $x = k \xi$.}
\label{fig:Psiplot1}
\end{figure}

In one dimension the static spin-spin correlation function $G ( \bd{k} )$ of a
ferromagnet assumes the Ornstein-Zernike form (\ref{eq:oz_static})
for small wavevectors~\cite{Takahashi86, Takahashi87},
where the uniform susceptibility $\chi$ and the correlation length
$\xi$ are related by
 \begin{equation}
 \chi = \rho^{-1} \xi^2 \propto 1/ T^2.
 \end{equation}
Here $\rho$ is the renormalized spin stiffness.
Explicit expressions for $\chi$ and $\xi$ have been obtained
by Takahashi within his modified 
spin-wave theory \cite{Takahashi86,Takahashi87}.
For a spin-$S$ Heisenberg chain with 
ferromagnetic 
nearest-neighbor coupling $J > 0$
Takahashi's result for the correlation length is \cite{Takahashi86,Takahashi87}
 \begin{equation}
\frac{ \xi}{a} =  \frac{J S^2}{T}.
 \label{eq:xiTaka}
 \end{equation}
A renormalization group calculation augmented by a Monte-Carlo simulation for $S = 1/2$ \cite{Kopietz89} suggests that for small $S$ the prefactor
in Eq.~(\ref{eq:xiTaka}) may not be correct, but this is irrelevant for our purpose.
Substituting the Ornstein-Zernike form (\ref {eq:oz_static}) 
for  the static spin-spin correlation function into the modified
kernel (\ref{eq:Vmod}) we find 
  \begin{equation}
  {V}_{\ast} ( \bd{k} , \bd{q} ) = \frac{T \rho  }{4} \frac{[1 + (k \xi )^2] 
[ k^2 + 2 \bd{k} \cdot \bd{q} ]^2 \xi^2 }{
 [ 1 + ( q \xi )^2 ][  1 + ( \bd{k} + \bd{q} )^2 \xi^2 ] }
 + ( \bd{k} \rightarrow - \bd{k} ),
 \label{eq:Vkernel}
 \end{equation}
which has the same form as the kernel $V ( \bd{k} , \bd{q} )$
in Eq.~(\ref{eq:Vkernel0}) with the bare spin stiffness $\rho_0$ replaced by the
renormalized one. The integral equation (\ref{eq:Deltaint}) for the dissipation energy has therefore
the same form as Eq.~(\ref{eq:Deltapara}) with  $\rho_0 $ replaced by $\rho$.
As a result, the dissipation energy 
$\Delta ( {\bd{k}} , i \omega )$ can again be written
in the scaling form, 
 \begin{equation}
  \Delta ( \bd{k} , i \omega )   = \tau^{-1} A ( k \xi, i \omega \tau ),
 \label{eq:Deltascale2}
 \end{equation}
where the scaling function $A ( x , iy )$ 
is the same as in Eq.~(\ref{eq:integralg})
and the characteristic time scale $\tau$ is in $d$ dimensions given by
 \begin{equation}
 \tau = \frac{ ( \xi /a )^{1 + d/2}}{\sqrt{T \rho /(2 a^2)}}.
 \end{equation}
Note that in one dimension at low temperatures
 \begin{equation}
 \tau  = \sqrt{\frac{2\chi(\xi/a)}{T}} \propto \frac{ \xi^{3/2}}{\sqrt{T}} \propto \xi^2,
 \end{equation}
so that the dynamical exponent in one dimension is $z=2$; the result $z = 1 + d/2$ based on dimensional analysis is not valid
in this case due to the vanishing of the critical temperature.
From Eq.~(\ref{eq:integralg}) we see that in $d=1$ the scaling function $A (x , iy )$ satisfies the
integral equation
 \begin{eqnarray}
 A ( x , i y ) & = & [1 + x^2]\int_{ - \infty}^{\infty}  \frac{ d r}{  2 \pi }  
 \frac{  ( x^2 + 2 x
 r  )^2}{ ( 1 + r^{ 2}) [ 1 + ( x + r )^2 ]}
 \nonumber
 \\
 & & \hspace{15mm} \times
 \frac{1}{A (r , i y ) + | y |}.
 \label{eq:integralA1}
 \end{eqnarray}
In Figs.~\ref{fig:Aplot1} and \ref{fig:Phiplot1} we show
our numerical results for the scaling function $A ( x, iy )$  and the
corresponding scaling function $\Phi ( x , y )$ of the dynamic structure factor
defined via Eqs.~(\ref{eq:Sscale1}) and (\ref{eq:phiscale}) in one dimension.
While results from mode-coupling theory and perturbative  renormalization group
calculations  based on the $\epsilon = 6-d$ expansion are not available for $d=1$ (the
extrapolation of a low-order expansion  in powers of $\epsilon$ to the case $\epsilon =5$ 
is obviously problematic)
the dynamic structure factor of a ferromagnetic Heisenberg chain has been calculated by 
Takahashi \cite{Takahashi90} using his  modified spin-wave theory \cite{Takahashi86,Takahashi87}.
Note that Takahashi's result for the uniform susceptibility is \cite{Takahashi86, Takahashi87} 
\begin{equation}
\chi = \frac{2JS^4}{3T^2},
 \label{eq:chiTaka}
\end{equation}
so that together with $\xi$ from Eq. \eqref{eq:xiTaka} our estimate for the relaxation time is 
\begin{equation}
\tau = \frac{2(\xi/a)^2}{\sqrt{3}JS},
\end{equation}
which is larger by a factor of $2/\sqrt{3} \approx 1.15$ compared to the expression given by Takahashi in \cite{Takahashi90}.
At the first sight, the spectral line-shape shown in Fig.~\ref{fig:Phiplot1} seems to resemble the line-shapes
obtained by Takahashi \cite{Takahashi90}.
In particular, for sufficiently large $x = k \xi$  the peak positions  disperse 
as $ x^2 \propto k^2$,
which can be identified with the dispersion of overdamped ferromagnetic magnons in the paramagnetic regime. 
To see this more clearly, it is useful to plot the scaling function of the dynamic structure factor
as a function of the scaling variable $ \nu = \omega \tau / ( k \xi )^2$ where we have used $z=2$.
Defining the scaling function $\Psi ( x , \nu )$ as in Eq.~(\ref{eq:SPsi}) with characteristic frequency
$\omega_k = ( k \xi )^2 / \tau $ we have 
 \begin{equation}
 \Psi ( x , \nu ) = x^2 \Phi ( x , \nu x^2 ),
 \label{eq:PsiPhi2}
 \end{equation}
which is Eq.~(\ref{eq:PsiPhi}) for $z=2$. A plot of the scaling function $\Psi ( x , \nu )$ as a function of $\nu$ for different values of
$x$ is shown in Fig.~\ref{fig:Psiplot1}. For large $x$ the curves collapse, proving therefore the $k^2$-dispersion of overdamped paramagnons. 
Note that our theory predicts
that the asymptotic behavior of the peak width has the same order as the dispersion.
This  disagrees with  the prediction of Takahashi's modified spin-wave 
theory \cite{Takahashi90}, who finds that the ratio of width to peak position vanishes like $x^{-1}$ for $x \rightarrow \infty$, implying increasingly well-defined spin-wave excitations for large
$x = k \xi$.
This discrepancy might be due to the fact  that 
Takahashi's modified spin-wave theory does not take  
spin-wave scattering into account and therefore underestimates the
decay rate of spin waves \cite{Reiter93,Takahashi93}.

Finally, let us take a closer look at 
the  small-frequency behavior of the dynamic structure factor in one dimension, 
which for fixed $x$ is determined
by the singularity of the scaling function $A ( x , iy )$ for $y \rightarrow 0$. 
From the integral equation (\ref{eq:integralA1}) we find for $y, x \ll 1$
 \begin{equation}
 A ( x, iy ) \sim A_1 x^2 | y |^{1/5} + A_2 \frac{x^4}{ | y |^{3/5}},  
 \label{eq:A12one}
 \end{equation}
where $A_1$ and $A_2$ are numerical constants. For large momenta $x \gg 1$ the dissipation energy consists of a sole term $\sim x^4 |y|^{-3/5}$.
Eq.~(\ref{eq:A12one}) is the one-dimensional analogue of the
corresponding expression 
(\ref{eq:expanded_diss_energy}) in three dimensions. 
The non-analytic frequency dependence is responsible for the pseudogap 
in the spectral line-shape  at small frequencies visible in Figs.~\ref{fig:Phiplot1} and \ref{fig:Psiplot1} and implies that for $\omega \rightarrow 0$
the dynamic structure factor vanishes as $ S ( \bd{k} , \omega )  \propto \omega^{3/5}$, which agrees with the behavior in three dimensions, 
see Eq.~(\ref{eq:Ssmall}).  As a consequence, $S ( \bd{k} , \omega )$
exhibits again a pseudogap for small frequencies and a
peak  at finite frequencies even in the hydrodynamic regime $x \ll 1$.
Our result excludes normal  diffusive behavior in one dimension. 
This is  in disagreement
with the prediction of modified spin-wave theory,  where 
for $x \lesssim  1$ the line-shape 
exhibits a single  elastic zero-frequency peak and for larger $x$ one observes a 
flat minimum for small $\omega$ with $S(\bm{k},0) > 0$, see Fig.~1 of Ref.~[\onlinecite{Takahashi90}].

In principle the dynamic structure factor
$S ( \bd{k} , \omega )$ of the spin $1/2$  Heisenberg chain with
nearest neighbor  coupling can be obtained from the 
thermodynamic Bethe-Ansatz although the explicit evaluation of the
relevant matrix elements can only be carried out approximately via  
a form-factor expansion.
While the dynamic structure factor of spin chains with  antiferromagnetic coupling has been 
discussed in the literature~[\onlinecite{Karbach97,Caux08,Mourigal13}], 
we have not been able to find published
Bethe-ansatz results for the dynamic structure factor of the
ferromagnetic spin chain.
Note that for a ferromagnetic chain the groundstate is non-degenerate so that
at low temperatures the density of excitations is expected to be small,
which makes the suppression of spectral weight 
for $\omega \rightarrow 0$ plausible. 
The pseudogap scenario is also supported by calculations in the limit of infinite temperature
 which predict anomalous diffusion with $\Delta(\bm{k},i\omega)$ diverging as $\omega^{-1/3}$ and are believed to converge at least for integrable 
spin chains~\cite{Dupont20,Bulchandani21}.

\subsection{Two dimensions}

\begin{figure}[tb]
\begin{center}
 \centering
\vspace{7mm}
\includegraphics[width=0.45\textwidth]{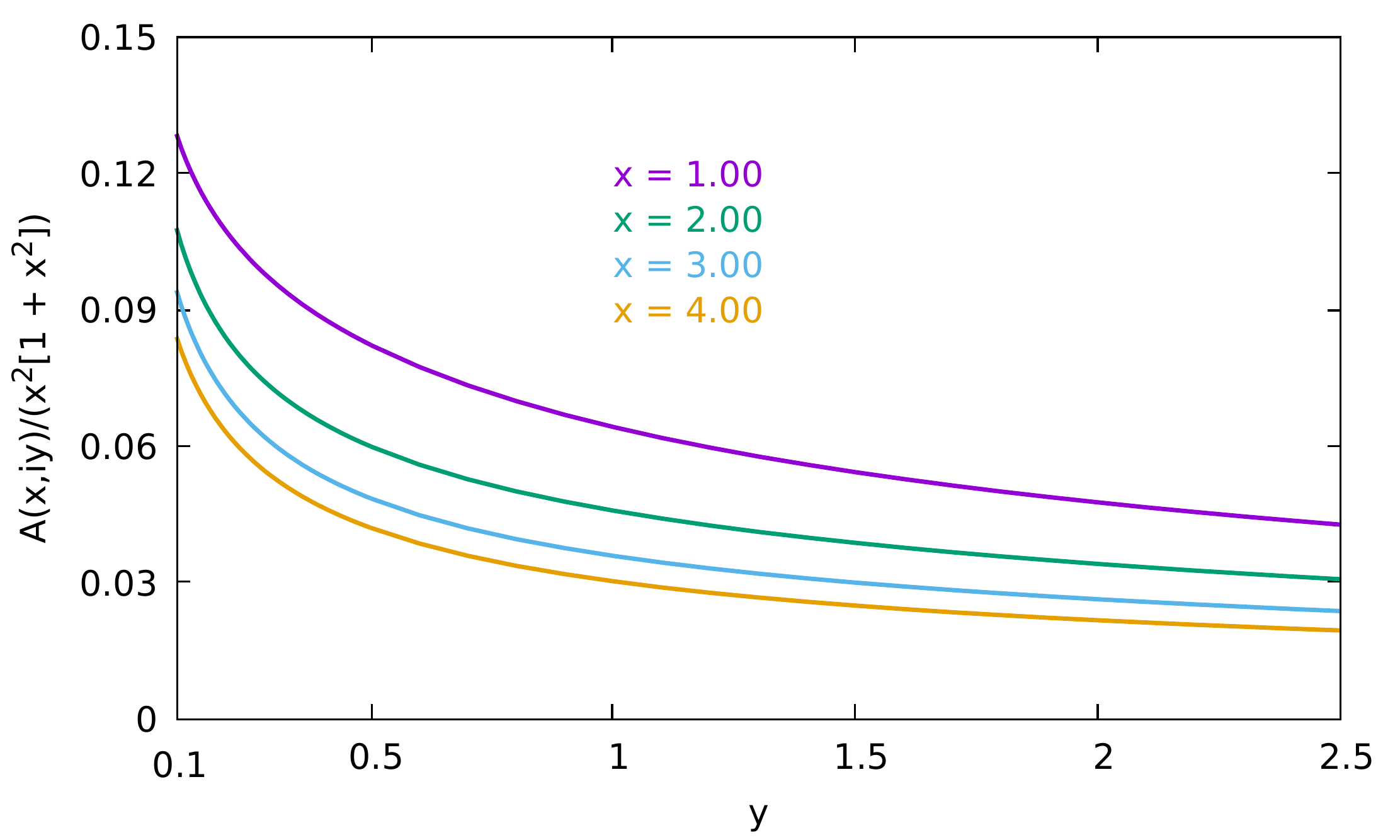} 
  \end{center}
  \caption{Frequency dependence of the scaling function $ A (x,iy)$ 
of the dissipation energy $\Delta ( \bm{k} , i \omega )$
in two dimensions obtained from the numerical solution of the
integral equation (\ref{eq:intA2}). We plot again
$A( x , iy )/ ( x^2[1 + x^2])$ to remove the dominant  $x$-dependence.
}
\label{fig:Aplot2}
\end{figure}

\begin{figure}[tb]
\begin{center}
 \centering
\vspace{7mm}
\includegraphics[width=0.45\textwidth]{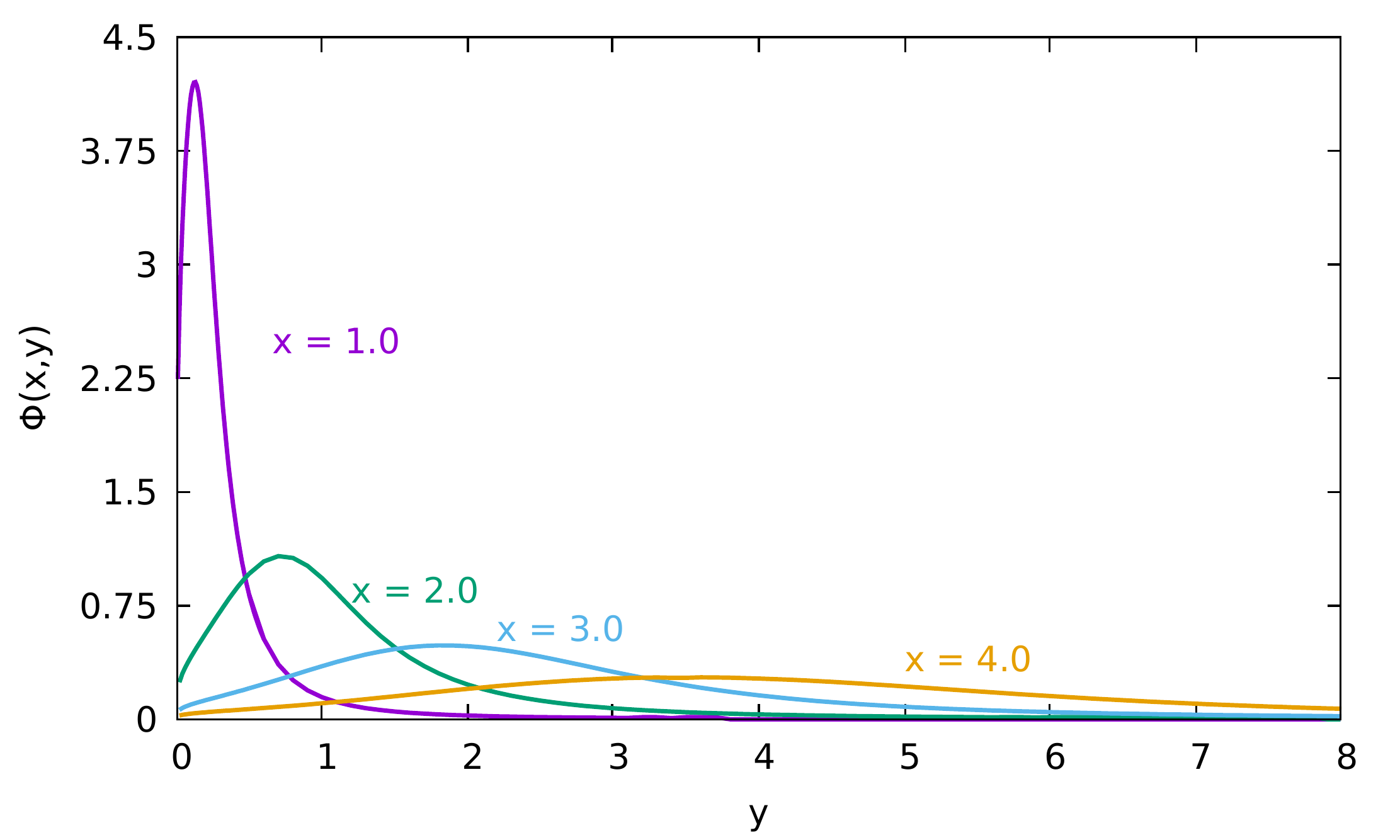}
   \end{center}
  \caption{Frequency dependence of the scaling function $\Phi(x,y)$ of the dynamic structure factor in two dimensions 
for different values of $x$, obtained by inserting  the numerical solution
of Eq.~(\ref{eq:intA2}) into the relation (\ref{eq:phiscale}).
}
\label{fig:Phiplot2}
\end{figure}

\begin{figure}[tb]
\begin{center}
 \centering
\vspace{7mm}
\includegraphics[width=0.45\textwidth]{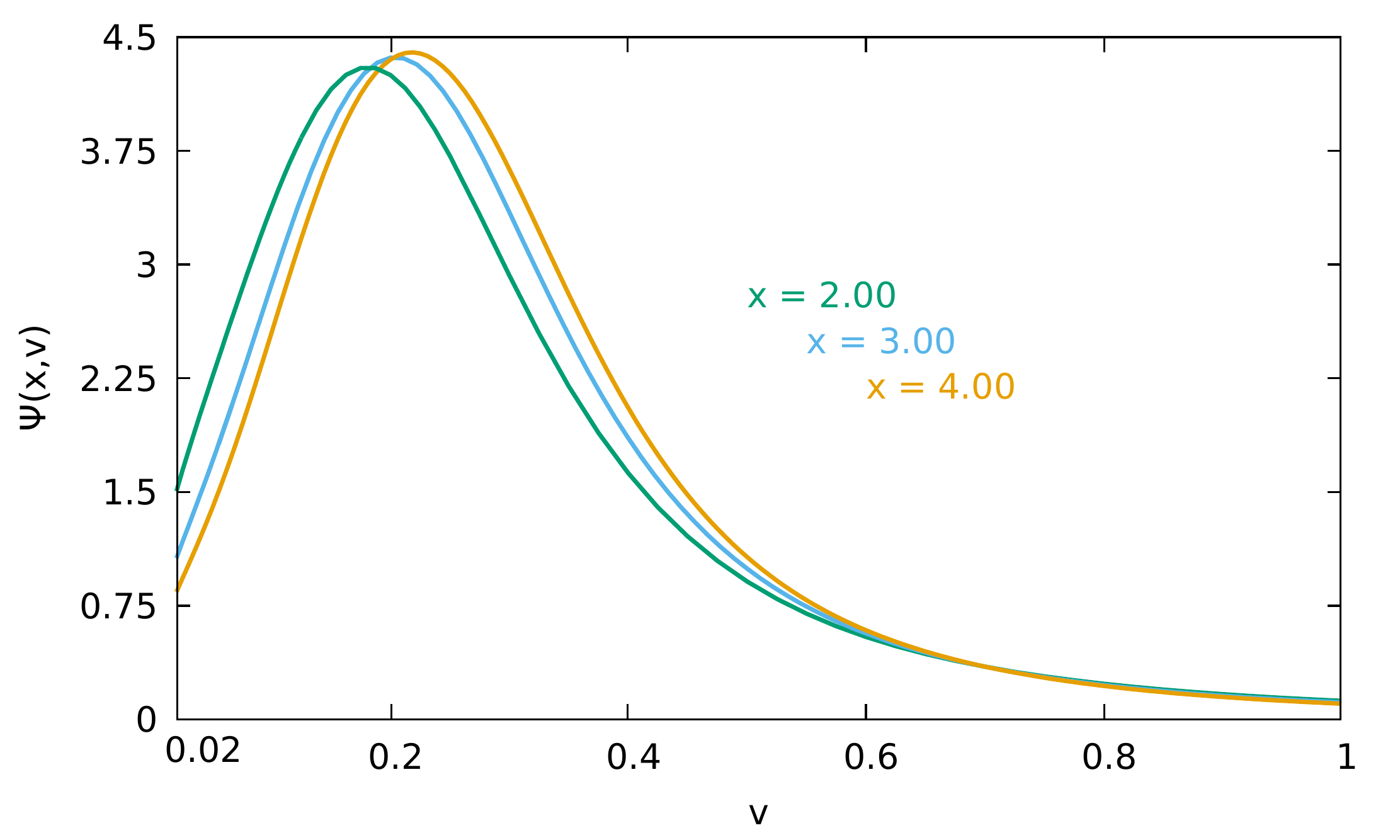}
   \end{center}
  \caption{Frequency dependence of the scaling function $\Psi(x,\nu)$
of the dynamic structure factor in two dimensions 
for different values of $x$. This function is 
defined via Eq.~(\ref{eq:SPsi}) and is related to $\Phi ( x, y )$
as in Eq.~(\ref{eq:PsiPhi2}).
}
\label{fig:Psiplot2}
\end{figure}

Finally, let us briefly discuss the case of two dimensions where
the Ornstein-Zernike ansatz does not correctly describe
the static susceptibility $G(\bm{k})$. For small momenta $ k  a \ll 1$
the result of   modified spin-wave theory is \cite{Takahashi90} 
\begin{equation}
G(\bm{k}) = \chi g(k\xi), 
 \label{eq:scaleG}
\end{equation}
with static scaling function
\begin{equation}
g(x) = \frac{\ln(x + \sqrt{x^2 + 1})}{x\sqrt{x^2 + 1}}.
\end{equation}
The correlation length $\xi$ and the uniform susceptibility are exponentially large at low 
temperatures \cite{Takahashi86,Takahashi87,Kopietz89b}
 \begin{eqnarray}
 \frac{\xi}{a} & \sim & C_\xi e^{ \alpha / T },
 \\
 \chi & \sim &  C_{\chi} e^{ 2 \alpha /T },
 \end{eqnarray}
where for nearest neighbor coupling $J$ on a square lattice $\alpha = 2 \pi J S^2$.
Modified spin-wave theory \cite{Takahashi86, Takahashi87} gives in this case $C_{\xi} = \sqrt{ JS/T}$  and $C_{\chi} = 1/(12\pi J S )$
so that at low temperatures
 \begin{equation}
 \frac{\chi}{  T \xi^2} = {\rm const.}
 \label{eq:chixi2}
 \end{equation}
A  more accurate two-loop 
RG calculation \cite{Kopietz89b} actually leads to a different  temperature-dependence of $C_\xi$ and $C_{\chi}$
but does not modify the relation (\ref{eq:chixi2}).
Substituting the scaling from (\ref{eq:scaleG}) for the static spin-spin correlation function 
into the kernel $\tilde{V} ( \bd{k} , \bd{q} )$ given in Eq.~(\ref{eq:Vmod})
we find that the dissipation energy
assumes again the dynamic scaling form (\ref{eq:Deltascale2}) where the
scaling function $A ( x , i y )$ now satisfies the integral equation
 \begin{eqnarray}
 A ( x , iy ) & = & g^{-1}(x) \int \frac{ d^2 r}{ ( 2 \pi )^2}
 g (r ) g ( |  \bd{x} + \bd{r} | )
 \nonumber 
 \\
 & & \times \frac{ \left[ g^{-1} ( r ) - g^{-1} ( | \bd{x} + \bd{r} | )
 \right]^2}{ A ( r , iy ) + | y | }.
 \label{eq:intA2}
 \end{eqnarray}
The characteristic time-scale $\tau$ is in two dimensions given by
 \begin{equation}
 \tau = \sqrt{ \frac{2\chi}{T} } \frac{\xi}{a} = 
  \sqrt{ \frac{2\chi a^2 }{T \xi^2} } \left(  \frac{\xi}{a} \right)^2.
 \end{equation}
Using the fact that according to Eq.~(\ref{eq:chixi2})
the combination $\chi /( T \xi^2 )$ approaches a temperature-independent constant, we conclude that in two dimensions $\tau \propto \xi^2$, implying $z =2$.
For nearest neighbor interaction $J > 0$
we may estimate the constant in Eq.~(\ref{eq:chixi2})
by inserting for $\xi$ and $\chi$  the results obtained
by Takahashi within modified spin-wave theory \cite{Takahashi87, Takahashi86} which gives
 \begin{equation}
\tau 
\approx \frac{1}{ \sqrt{6 \pi} J S }
 \left( \frac{\xi }{ a } \right)^2.
 \end{equation}
This is a factor of  $ ( 6\pi)^{-1/2} \approx 0.23$ smaller than the value given
in Ref.~[\onlinecite{Takahashi90}]. 
In Fig.~\ref{fig:Aplot2} we show our result for $A ( x , iy )$ as a function of the
dimensionless frequency $y$ obtained from the numerical solution of
Eq.~(\ref{eq:intA2}).
The corresponding scaling function $ \Phi ( x , y )$
of the dynamic structure factor
is shown in Fig.~\ref{fig:Phiplot2}.
In order to exhibit the dispersion of the peak, we plot
 in Fig.~\ref{fig:Psiplot2}
 the scaling function $\Psi ( x , \nu )$ using $\nu = y / x^2 =
 \omega \tau / ( k \xi )^2$ as independent variable.
The qualitative behavior of the scaling functions is similar to the one-dimensional case:
the dynamic structure factor exhibits a pseudogap for frequencies
$\omega \lesssim \omega_k = ( k \xi )^2 / \tau$ and a peak which disperses as
$k^2$ for $k \gtrsim \xi^{-1}$ and can be interpreted as an overdamped paramagnon.
Note that the dynamic scaling functions $\Phi(x,y)$ and $\Psi(x,\nu)$ describe only the leading frequency dependence of $S(\bm{k},\omega)$ as long as $\omega \ll T$, so that the detailed-balance factor can still be approximated by the classical expression given in Eq.~\eqref{eq:classflucdiss}. Since this regime shrinks with decreasing $T$, the zero-temperature limit is not covered by these calculations.

\section{Summary and Conclusions}

In this work we have investigated the spin dynamics of
quantum Heisenberg ferromagnets with isotropic exchange interactions
in the paramagnetic phase close to  the critical temperature
by solving the integral equation for the spin-spin correlation function
derived in Ref.~[\onlinecite{Tarasevych21}] within the framework of
a novel functional renormalization group approach to quantum spin systems  \cite{Krieg19}.
Our results  are consistent with the validity of the dynamic scaling 
hypothesis \cite{Halperin69,Hohenberg77} and we obtain explicit expressions for the
scaling functions describing the  line-shape of the dynamic structure factor
$S ( \bd{k} , \omega )$. 
Note that dynamic scaling on its own does 
not make any statements about the actual line-shape close to  the critical point.

In three dimensions our approach  reproduces the
known behavior of the spin-diffusion coefficient \cite{Kawasaki67}
in the hydrodynamic regime. However, we have also detected corrections to hydrodynamics
associated with  non-analytic corrections to diffusion, 
which partially have been 
discussed previously  in Refs.~[\onlinecite{Borckmans77,Fogedby78, Pomeau75}].
The form of these corrections implies  a non-Lorentzian line-shape of $S(\bm{k},\omega)$ down to $\omega = 0$ which  deviates for small frequencies 
from results obtained in mode-coupling theory. Such non-hydrodynamic terms play 
only a role if one considers arbitrarily long times $t$ at fixed momentum $\bd{k}$. 
In the strict hydrodynamic limit where $\bd{k}  \rightarrow 0$ with  constant 
$k^2t$, which is considered in most calculations, 
these corrections are negligible.

Precisely at the critical point we have obtained  the  dynamic structure factor in the
scaling form $S ( \bd{k} , \omega ) = T_c G ( \bd{k} )(\pi \omega_k)^{-1} \Psi_c ( \omega / 
 \omega_k )$, where $\omega_k \propto k^{z}$ with $z = 5/2$ in three dimensons,
see Eq.~(\ref{eq:Psicdef}). For large $\nu = \omega / \omega_k$
the critical scaling function $\Psi_c ( \nu )$ exhibits a non-Lorentzian decay proportional to
$ \nu^{-13/5}$, in agreement with previous calculations using either mode-coupling 
theory~\cite{Wegner68} or the asymptotic renormalization group based
on the extrapolation of the expansion in powers of
$\epsilon = 6-d$ to the physically relevant case of three dimensions \cite{Bhattacharjee81,Folk85}.
On the other hand, for small $\nu$ our result for  
$\Psi_c ( \nu )$ drastically differs from
previous approximate calculations \cite{Wegner68,Hubbard71,Bhattacharjee81, Folk85,Frey94,Folk06} which predict a single peak at vanishing frequency which is however broader
than the Lorentzian in the hydrodynamic regime. In contrast,
 we find that $\Psi_c ( \nu )$ exhibits a finite-frequency peak at $\nu \approx 0.2$
and vanishes as $\nu^{3/5}$ for $\nu \rightarrow 0$, leading to a pseudogap in the critical
line-shape for small frequencies.

Unfortunately, 
available numerical simulations \cite{Chen94,Tao05} 
do not have sufficient accuracy to resolve this discrepancy. Moreover, as discussed
in Sec.~\ref{sec:experiments},
neutron scattering experiments  \cite{Boeni86,Boeni87,Boeni88} probing the critical
spin-dynamics in the  Heisenberg ferromagnets EuO and EuS
 are somewhat inconclusive due to a limited energy resolution, preventing us from checking directly on the existence of the aforementioned low-frequency feature. 
Taking the finite energy resolution of the experiments into account,
the line-shapes obtained within our approach agree reasonably well with the
experiments. However, this is true even more for the critical line-shapes
obtained within asymptotic renormalization group theory \cite{Bhattacharjee81,Folk85}, so that
we have to admit that, at least in three dimensions, 
the pseudogap of the critical line-shape obtained within our approach might be an unphysical feature of our method.

It is interesting to see how the algebraic singularity of
$S ( \bd{k} , \omega )$ for small $\omega$ at the critical point evolves with the dimensionality $d$ of the system.
By directly taking the limit $\xi 
 \rightarrow \infty$ in the
integral equation (\ref{eq:Deltapara}) for the dissipation energy we find that in 
dimensions $ 2 < d < 6$ the low-frequency behavior of the 
critical scaling function $B_c ( i \nu )$ defined  in Eq.~(\ref{eq:Bcdef}) is
 \begin{equation}
 B_c ( i \nu ) \propto | \nu |^{1 - 4/z}, \; \; \; z = 1 + d/2.
 \end{equation}
Eqs.~(\ref{eq:Psicdef}) and (\ref{eq:PsiBscalecrit}) then  imply
that for $\omega \rightarrow 0$  the critical dynamic structure factor 
in dimensions $2 < d < 6$ vanishes as
 \begin{equation}
 S ( \bd{k} \neq 0 , \omega ) \propto \omega^{-1 + 4/z} = 
\omega^{\frac{6-d}{2 + d}}, \; \; \; \mbox{for $\omega \rightarrow 0$.}
 \label{eq:Sasymd}
 \end{equation}
The pseudogap therefore disappears for  $d \geq 6$ where
$S ( \bd{k} \neq 0, \omega )$ approaches a finite limit for $\omega 
 \rightarrow 0$. On the other hand, the pseudogap  
becomes more pronounced in lower dimensions, so that we expect
that in this case our approximate method give  a better description of
the relevant fluctuations responsible for the suppression of spectral weight for small frequencies.  Of course, in $d \leq 2$ our 
estimate (\ref{eq:Sasymd}) is not valid because it is based on the assumption of
a finite critical temperature; however,  in  Sec.~\ref{sec:reduced} we have shown that
also in
one and two dimensions the low-temperature behavior of the spectral line shape
exhibits a pseudogap for small frequencies.
Interestingly, we have found that both in $d=1$ and $d=2$  the 
relevant characteristic energy $\omega_k$ scales as $k^2$ at low temperatures, which 
agrees with the dispersion of ferromagnetic magnons at $T=0$.
The pseudogap and the finite-frequency maximum in the line-shape of $S ( \bd{k} , \omega )$ at 
finite $T$ can therefore be associated with overdamped magnons in the paramagnetic phase.
Note that ordinary diffusive behavior would lead to a zero-frequency maximum
in the spectral line shape, so that we conclude that in $d=1$ and $d=2$ 
ordinary diffusion does not emerge  even for hydrodynamic momenta $k  \ll 1/ \xi$.

We conclude that the problem of determining  the low-frequency behavior of the line-shape
of the dynamic structure factor close to the critical point
of an isotropic Heisenberg ferromagnet is still not completely solved. 
Available analytical calculations 
all rely on some kind of uncontrolled approximation (including our approach).
Moreover,
neither numerical simulations nor available experiments have so far produced
data with sufficiently high resolution.  
In one dimension, the
line-shape of the dynamic structure factor  of the spin $1/2$ Heisenberg chain 
can in principle be calculated in a controlled way 
using the thermodynamic
Bethe ansatz, but  the case of a ferromagnetic coupling has so far not been analyzed.

\section*{Acknowledgements}
We thank Oleksandr Tsyplyatyev for his comments concerning the
Bethe ansatz for the ferromagnetic Heisenberg chain.
This work was completed during a sabbatical stay of P.K. at the
Department of Physics and Astronomy at the University of California, Irvine.
P.K. would like to thank Sasha Chernyshev
for his hospitality.
We also thank the  Deutsche Forschungsgemeinschaft (DFG) for financial support 
through project KO 1442/10-1.

\begin{appendix}

\section*{APPENDIX: Spin FRG with classical-quantum
decomposition}

\setcounter{equation}{0}
\renewcommand{\theequation}{A\arabic{equation}}

To make this work self-contained, we outline here the
derivation of our integral equation (\ref{eq:Deltaint}) for the dissipation energy $\Delta ( \bd{k} , i \omega)$ 
defined via Eq.~(\ref{eq:GMatsubara})
within our recently developed spin FRG approach. For more details we refer the reader to Ref.~[\onlinecite{Tarasevych21}].
To begin with, let us consider a general spin-rotationally invariant
Heisenberg Hamiltonian
  \begin{equation}
 {\cal{H}} =  \frac{1}{2} \sum_{ij} {J}_{ij} \bd{S}_i \cdot \bd{S}_j,
 \label{eq:Heisenberg}
 \end{equation}
where $J_{ij} = J ( \bd{R}_i - \bd{R}_j )$ are arbitrary exchange couplings connecting spin-$S$
operators $\bd{S}_i$ localized at the sites $\bd{R}_i$ of 
a $d$-dimensional Bravais lattice.
Choosing periodic boundary conditions we may expand
the exchange interactions  in a Fourier series,
 \begin{equation}
 J_{ij} = \frac{1}{N} \sum_{\bd{k} } e^{ i \bd{k} \cdot ( \bd{R}_i - \bd{R}_j ) }
 J ( \bd{k} ),
 \end{equation}
where the $\bd{k}$-sum is over the first Brillouin zone. 

Following Ref.~[\onlinecite{Krieg19}] we now replace the $J_{ij}$ by
deformed exchange couplings $J_{ij} \rightarrow J^{\Lambda}_{ij}$ depending continuously on a deformation parameter $\Lambda$ and follow the
evolution of the generating functional  ${\cal{G}}_{\Lambda} [ \bd{h} ] $
of the imaginary-time ordered connected  
spin correlation functions when  $\Lambda$ evolves from some initial value
$\Lambda = \Lambda_0$ down to $\Lambda =0$ where $J_{ij}^{\Lambda =0}
 = J_{ij}$. The generating functional  ${\cal{G}}_{\Lambda} [ \bd{h} ] $ is defined by
 \begin{eqnarray}
 & & {\cal{G}}_{\Lambda} [ \bd{h} ]  =
 \nonumber
 \\
 &  & 
\ln {\rm Tr}
 \left[  {\cal{T}} e^{ \int_0^{\beta} d \tau 
 \left[
\sum_i \bd{h}_i ( \tau )
 \cdot \bd{S}_i ( \tau ) - \frac{1}{2} \sum_{ij} J^{\Lambda}_{ ij}
 \bd{S}_i ( \tau ) \cdot \bd{S}_j ( \tau ) \right]} \right],
 \hspace{7mm}
 \label{eq:Gcondef}
 \end{eqnarray}
where
$\bd{h}_i ( \tau )$ is  a fluctuating  magnetic source field,  ${\cal{T}}$ denotes time-ordering
in imaginary time, and the imaginary-time label $\tau$ of the spin operators 
$ \bd{S}_i ( \tau )$ keeps track of the time-ordering. 
By taking a partial derivative of Eq.~(\ref{eq:Gcondef})  
with respect to $\Lambda$ we obtain a formally exact closed 
flow equation \cite{Krieg19} for $\mathcal{G}_\Lambda[\bd{h}]$, 
implying an infinite hierarchy of integro-differential equations for the connected 
spin correlation functions.  Usually \cite{Wetterich93,Berges02,Pawlowski07,Kopietz10,Metzner12,Dupuis21}
one now introduces the (subtracted) Legendre transform  $\Gamma_{\Lambda} [ \bd{m} ]$ of
 ${\cal{G}}_{\Lambda} [ \bd{h} ]$ which depends on the magnetization $\bd{m}$, generates the one-particle irreducible spin-vertices, and
satisfies the Wetterich equation \cite{Wetterich93}.
Unfortunately, in the case of the Heisenberg model this approach comes with a major intricacy, because for the exactly solvable initial condition $J^{\Lambda_0}_{ij} = 0$ of decoupled spins 
the two-spin correlation function for vanishing sources is zero for any finite Matsubara frequency due to spin conservation, thus implying in turn that the one-particle irreducible two-point vertex
diverges for finite Matsubara frequencies. 
As a consequence, the initial condition for all one-particle irreducible  vertices
involving finite frequencies is ill-defined and one cannot calculate a proper flow using this parametrization. 
In Refs.~[\onlinecite{Krieg19,Goll19,Goll20}]
several ways  to avoid this problem have been proposed, all of which
are based on the strategy of replacing $\mathcal{G}_\Lambda[\bd{h}]$
by a different type of generating functional whose Legendre transform
exists even if the exchange couplings are completely switched off 
while it still satisfies the Wetterich equation.
For example, one can start from the generating functional of connected correlation functions which are in addition amputated with respect to $J^\Lambda_{ij}$ \cite{Krieg19,Goll20} 
so that the corresponding vertices are irreducible  with respect to cutting a single interaction line.
The corresponding two-point vertices are identical with those considered in the 
diagrammatic approach to quantum spin systems developed many years ago 
by Vaks, Larkin, and Pikin \cite{Vaks68, Vaks68b},
who write the spin-spin correlation function in the form
 \begin{equation}
 G (\bm{k},i\omega) = \frac{\Pi (\bm{k},i\omega)}{1 +  J (\bm{k}) 
 \Pi(\bm{k},i\omega)}.
 \label{eq:GPi}
 \end{equation}
We refer to $\Pi (  \bd{k} , i \omega )$ as the interaction-irreducible 
dynamic susceptibility.

In the present work a slight  variation of this construction is more useful. 
Let us therefore note that in the static sector, 
where fluctuations with finite frequencies are  excluded, the Legendre transform of $\mathcal{G}_\Lambda[\bd{h}]$ can be written in terms of an analytic functional series expansion around $\bd{m} = 0$ even for vanishing 
exchange interaction. On the other hand, for finite frequencies
 the interaction-irreducible vertices are well-defined~\cite{Krieg19, Goll19}.
To take advantage of these properties,
we work within a hybrid approach which explicitly distinguishes between  static (i.e.
classical) and dynamic (i.e. quantum) fluctuations~\cite{Tarasevych21}.  
However, to make  this strategy work in the paramagnetic phase,  it is necessary
to define the amputation of the interaction with respect to the scale-dependent
subtracted exchange interaction
 \begin{equation}
 \tilde J_\Lambda(\bm{k}) = J_\Lambda(\bm{k}) + \Pi^{-1}_\Lambda(\bm{k},0) = G^{-1}_\Lambda(\bm{k}),
 \end{equation}
 which is just the inverse of the scale-dependent 
static susceptibility $G_\Lambda(\bm{k})$.
In analogy with Eq.~(\ref{eq:GPi})
we therefore write the dynamic spin-spin correlation function in the form
 \begin{equation}
 G_\Lambda(\bm{k},i\omega) = \frac{\tilde \Pi_\Lambda(\bm{k},i\omega)}{1 + \tilde J_\Lambda(\bm{k})\tilde \Pi_\Lambda(\bm{k},i\omega)} = \frac{1}{ \tilde J_\Lambda(\bm{k}) + \tilde \Pi^{-1}_\Lambda(\bm{k},i\omega)}.
 \end{equation}
The subtracted scale-dependent generalized  susceptibility
 \begin{equation}
\tilde \Pi^{-1}_\Lambda(\bm{k},i\omega) = \Pi^{-1}_\Lambda(\bm{k},i\omega) - \Pi^{-1}_\Lambda(\bm{k},0)
 \end{equation}
is irreducible with respect to cutting a single subtracted interaction line.
The ergodic property~\cite{Alessio16,Chiba20,Tarasevych21}
 \begin{equation}
 \lim_{\omega \rightarrow 0} G(\bm{k} \neq 0,\omega) =  G(\bm{k}),
 \end{equation}
then implies
 \begin{equation}
 \tilde \Pi^{-1}_\Lambda(\bm{k} \neq 0,0) = \lim_{\omega \rightarrow 0} \tilde \Pi^{-1}_\Lambda(\bm{k} \neq 0,i\omega) = 0.
 \label{eq:pi_ergod}
 \end{equation}
Furthermore, spin-rotational invariance implies
 \begin{equation}
 \big[ \mathcal{H}, \sum_i \bm{S}_i \big] = 0,
 \end{equation}
so that the total spin $\sum_i \bd{S}_i$  is conserved and hence
 \begin{eqnarray}
 G_\Lambda(\bd{k}=0, i\omega \neq 0) & = &  0,
 \\
 \tilde \Pi_\Lambda(\bd{k}=0 , i\omega \neq 0) & = &  0.
 \label{eq:pi_cons}
 \end{eqnarray}
In order to arrive at a parametrization of the FRG flow
where the two-point vertex for vanishing  frequency can be
identified with  $\Sigma_\Lambda(\bm{k}) \equiv \Pi^{-1}_\Lambda(\bm{k}, 0)$
and for finite frequency with
 $\tilde \Pi_\Lambda(\bm{k},i\omega \neq 0)$ we introduce the auxiliary functional \cite{Tarasevych21}
  \begin{equation}
 {\cal{F}}_\Lambda [ {\bd{h}}^c , {\bd{s}}^q ] = {\cal{G}}_{\Lambda} [  \bd{h}^c, \bd{h}^q = - \tilde{\mathbf{J}}_{\Lambda} \bd{s}^q ]
 - \frac{1}{2} ( \bd{s}^q, \tilde{\mathbf{J}}_{\Lambda} \bd{s}^q ),
 \label{eq:Fhybrid}
 \end{equation}
where the matrix elements of the matrix $\tilde{\mathbf{J}}_{\Lambda}$ 
are the subtracted exchange couplings $ \tilde{J}_{\Lambda} ( \bd{k} )$
and we have decomposed 
the magnetic source field into classical and quantum components, i.e.,
 \begin{equation}
 \bd{h}_{i , \omega } = \beta \delta_{\omega,0} \bd{h}_i^c + ( 1 - \delta_{\omega , 0} ) \bd{h}_{i, \omega}^q.
 \end{equation}
Differentiation of the above  auxiliary functional  ${\cal{F}}_\Lambda [ {\bd{h}}^c , {\bd{s}}^q ]$
with respect to the sources 
generates connected spin correlation functions with the additional properties that
in the quantum sector  external interaction lines are amputated. 
The corresponding two-point function at finite frequencies can then be interpreted as an effective subtracted exchange interaction, while higher order correlation functions can be obtained
from their connected counterparts by multiplying the quantum legs by factors of $ -\tilde{\mathbf{J}}_{\Lambda}$.  Our hybrid functional with the desired properties is now given by the subtracted Legendre transform of the above auxiliary functional 
${\cal{F}}_\Lambda [ {\bd{h}}^c , {\bd{s}}^q ]$, 
 \begin{eqnarray}
 \Gamma_{\Lambda} [ {\bd{m}}^c , {\bd{\eta}}^q ] & = & ( {\bd{m}}^c, {\bd{h}}^c ) + 
 ( {\bd{\eta}}^q , {\bd{s}}^q ) 
 - {\cal{F}}_\Lambda [ {\bd{h}}^c , {\bd{s}}^q ] 
 \nonumber
 \\
 &- &  \frac{1}{2} ( {\bd{m}}^c , \mathbf{R}_\Lambda^c  {\bd{m}}^c ) - 
 \frac{1}{2} ( \bd{\eta}^q , \mathbf{R}_{\Lambda}^q  \bd{\eta}^q ),
 \hspace{7mm}
 \label{eq:Gammahybrid}
 \end{eqnarray}
 where the classical (zero-frequency)  
field $\bd{m}^c$ is the first derivative of $\mathcal{F}_\Lambda$ with respect to the classical source field $\bd{h}^c$
while the quantum field $\bd{\eta}^q$ is the first derivative of 
$\mathcal{F}_\Lambda$ with respect to the quantum source $\bd{s}^q$.
The regulators $\mathbf{R}^{c}_{\Lambda}$ 
and  $\mathbf{R}^{q}_{\Lambda}$ parametrize the deformed exchange coupling for
vanishing and finite frequencies.
In the classical limit $S \rightarrow \infty$, where the time-dependence of the spin operators can be neglected, 
the functional $\Gamma_{\Lambda} [ {\bd{m}}^c ,0]$
reduces to the average effective action $\Gamma_\Lambda[\bm{m}^c]$, 
which is the subtracted Legendre transform of the 
classical functional $\mathcal{G}_\Lambda[\bm{h}^c]$.  Note that even for finite spin lengths $S < \infty$ purely static vertices, where all frequencies are set to $0$, can be identified with the one-particle irreducible vertices 
generated by the subtracted Legendre transform of the quantum functional $\mathcal{G}_\Lambda[\bm{h}]$.
Following Ref.~[\onlinecite{Tarasevych21}]
it can now be shown that $\Gamma_\Lambda[\bm{m}^c, \bm{h}^q]$ satisfies a generalized Wetterich equation \cite{Tarasevych21}, which still has a residual tree-level term proportional to $\partial_\Lambda \Sigma_\Lambda$.
This tree term is generated
 because the $\Lambda$-derivatives of the bare and subtracted coupling do not coincide $\partial_\Lambda J_\Lambda \not= \partial_\Lambda \tilde J_\Lambda$. 
 Note that
the tree-term does not generate contributions to the flow of purely static vertices.
After a number of approximations described in detail in \cite{Tarasevych21} which are such that the ergodicity condition 
\eqref{eq:pi_ergod} and the constraint \eqref{eq:pi_cons} imposed by 
spin conservation are fulfilled, we obtain the following integral equation for the 
subtracted dynamic susceptibility,
\begin{equation}
\tilde \Pi(\bm{k},i\omega) = \frac{1}{ N \omega^2} \sum_{ \bd{q}}
    \frac{  \tilde{V} ( \bd{k} , \bd{q} ) }{   G ( \bd{q} ) + \tilde{\Pi} ( \bd{q} , i \omega )},
\end{equation}
where the dimensionless kernel $ \tilde{V} ( \bd{k} , \bd{q} )$ is defined by
 \begin{eqnarray}
 \tilde{V} ( \bd{k} , \bd{q} ) & = &   T \bigl[   {G} ( \bd{q} + \bd{k} ) Z ( \bd{q} , \bd{k} )
  +   {G} ( \bd{q} -  \bd{k}  ) Z ( \bd{q} , - \bd{k} ) 
 \nonumber
 \\
 & & \hspace{3mm}  -   2 {G} ( \bd{q} ) \bigr],
 \label{eq:kerneldef}
 \end{eqnarray}
and the vertex renormalization factor $Z ( \bd{q} , \bd{k})$
is defined in Eq.~(\ref{eq:Zdef}).
 Note that the kernel is completely determined by static quantities, i.e. the static susceptibility $G(\bm{q})$. Introducing the dissipation energy
$\Delta ( \bd{k} , i \omega )$ via
 \begin{equation}
 \tilde{\Pi} ( \bd{k} , i \omega ) = G ( \bd{k} ) \frac{ \Delta ( \bd{k} , i \omega ) }{ | \omega | },
 \end{equation}
we finally 
arrive at the integral equation (\ref{eq:Deltaint}) given in Sec.~\ref{sec:intro}.

\end{appendix}


\begin{thebibliography}{99}

\bibitem{VanHove54}
L. Van Hove, {\it{Time-Dependent Correlations between Spins and Neutron Scattering in Ferromagnetic Crystals}},
Phys. Rev. {\bf{95}}, 1374 (1954).
%
\bibitem{Kawasaki67}
K. Kawasaki, { \it{Anomalous spin diffusion in ferromagnetic spin systems}}, J. Phys. Chem. Solids {\bf{28}}, 1277 (1967).
%

\bibitem{Wegner68}
F. Wegner, {\it{On the Heisenberg model in the paramagnetic region and at the critical point}}, Z. Phys. {\bf{216}}, 433 (1968).
%
\bibitem{Halperin69}
B. I. Halperin and P. C. Hohenberg, {\it{Scaling Laws for Dynamic Critical Phenomena}}, Phys. Rev. {\bf{177}}, 952 (1969).
%
\bibitem{Hohenberg77}
P. C. Hohenberg and B. I. Halperin, {\it{Theory of dynamic critical phenomena}},
Rev. Mod. Phys. {\bf{49}}, 435 (1977).
%
\bibitem{Resibois70}
P. Resibois and C. Piette, {\it{Temperature Dependence of the Linewidth in Critical Spin Fluctuation}},
Phys. Rev. Lett. {\bf{24}}, 514 (1970).
%
\bibitem{Hubbard71}
J. Hubbard, {\it{Spin-correlation functions in the paramagnetic
phase of a Heisenberg ferromagnet}}, J. Phys. C: Solid State Phys. {\bf{4}}, 53 (1971).
%
\bibitem{Ma75}
S.-K. Ma and G. F. Mazenko,
{\it{Critical dynamics of ferromagnets in $6-\epsilon$ dimensions:
General discussion and detailed calculation}},
Phys. Rev. B {\bf{11}}, 4077 (1975).
%
\bibitem{Dohm76}
V. Dohm, {\it{Dynamical spin-spin correlation function of an isotropic ferromagnet
at $T_c$ in $6 - \epsilon$ dimensions}},
Solid State Commun. {\bf{20}}, 657 (1976).
%
\bibitem{Nolan77}
M. J. Nolan and G. F. Mazenko,
{\it{Dynamic structure factor for a ferromagnet in the scaling region to
first order in $\epsilon = 6-d$}},
Phys. Rev. B {\bf{15}}, 4471 (1977).
%
\bibitem{Borckmans77}
P. Borckmans, G. Dewel and D. Walgraef, {\it{Long time behaviour of correlation functions in Heisenberg paramagnets}}, Physica A, Volume {\bf{88}}, 2261 (1977).
%
\bibitem{Fogedby78}
H. C. Fogedby and A. P. Young, {\it{On correlations at long times in Heisenberg
paramagnets}}, J. Phys. C: Solid State Phys. {\bf{11}}, 527 (1978).
%
\bibitem{Bhattacharjee81}
J. K. Bhattacharjee and R. A. Ferrell, {\it{Dynamic scaling for the isotropic ferromagnet:
$\epsilon$ expansion to two-loop order}}, Phys. Rev. B {\bf{24}}, 6480 (1981).
%
\bibitem{Folk85}
R. Folk and H. Iro,  {\it{Paramagnetic neutron scattering and renormalization-group theory
for isotropic ferromagnets at $T_{c}$}}, Phys. Rev. B {\bf{32}}, 1880(R) (1985).
%
\bibitem{Frey89}
E. Frey, F. Schwabl, and S. Thoma, {\it{Shape functions of dipolar ferromagnets at and above the Curie point}}, Phys. Rev. {\bf{B}} 40, 7199 (1989).
%
\bibitem{Frey94}
E. Frey and F. Schwabl, {\it{Critical dynamics of magnets}}, Adv.  Phys., {\bf{43:5}}, 577 (1994).
%
\bibitem{Chen94}
K. Chen and D. P. Landau, {\it{Spin-dynamics study of the dynamic critical behavior of the three-dimensional classical Heisenberg ferromagnet}}, Phys. Rev. B {\bf{49}}, 3266 (1994). 
%
\bibitem{Tao05}
X. Tao, D. P. Landau, T. C. Schulthess, and G. M. Stocks, {\it{Spin Waves in Paramagnetic bcc Iron: Spin Dynamics Simulations}}, Phys. Rev. Lett. {\bf{95}}, 087207 (2005).
%
\bibitem{Folk06}
R. Folk and G. Moser, {\it{Critical dynamics: a field-theoretical approach}}, J. Phys. A: Math. Gen. {\bf{39}} R207 (2006).
%
\bibitem{Passell76}
L. Passell, O. W. Dietrich, and J. Als-Nielsen,
{\it{Neutron scattering from the Heisenberg ferromagnets EuO and EuS. I. The exchange interactions}},
Phys. Rev. B {\bf{14}}, 4897 (1976).
%
\bibitem{Wicksted84}
J. P. Wicksted, P. B\"{o}ni, and G. Shirane, {\it{Polarized-beam study of the paramagnetic scattering from bcc iron}},
Phys. Rev. B {\bf{30}},  3655 (1984).
%
\bibitem{Mezei86}
F. Mezei, {\it{Critical Dynamics in EuO at the Ferromagnetic Curie Point}}, Physica {\bf{B}}, 136, 417 (1986).
%
\bibitem{Boeni86}
P. B\"{o}ni and G. Shirane, {\it{Paramagnetic neutron scattering from the Heisenberg ferromagnet EuO}}, Phys. Rev. B {\bf{33}}, 3012 (1986).
%
\bibitem{Boeni87}
P. B\"{o}ni, M. E. Chen, and G. Shirane, {\it{Comparison of the critical magnetic scattering from the Heisenberg system EuO with renormalization-group theory}}, Phys. Rev. B {\bf{35}}, 8449 (1987).
%
\bibitem{Boeni88}
P. B\"{o}ni, G. Shirane, H. G. Bohn, and W. Zinn, {\it{Spin fluctuations in EuS above
$T_C$: Comparison with asymptotic
renormalization-group theory}}, J. Appl. Phys. {\bf{63}}, 3089 (1988).
%
\bibitem{Wilson72}
K. G. Wilson and M. E. Fisher, {\it{Critical Exponents in 3.99 Dimensions}},
Phys. Rev. Lett. {\bf{28}}, 240 (1972). 
%
\bibitem{Wetterich93}
C. Wetterich, {\it{Exact evolution equation for the effective potential}},
Phys. Lett. B {\bf{301}}, 90 (1993).
%
\bibitem{Berges02}
J. Berges, N. Tetradis, and C. Wetterich,
{\it{Non-perturbative renormalization flow in quantum field theory and statistical physics}},
Phys. Rep. {\bf{363}}, 223 (2002).
%
\bibitem{Pawlowski07}
J. M. Pawlowski, {\it{Aspects of the functional renormalisation group}},
Ann. Phys. {\bf{322}}, 2831 (2007).
%
\bibitem{Kopietz10}
P. Kopietz, L. Bartosch, and F. Sch\"{u}tz, {\it{Introduction to the Functional Renormalization
Group}}, (Springer, Berlin, 2010).
%
\bibitem{Metzner12}
W. Metzner, M. Salmhofer, C. Honerkamp, V. Meden, and K. Sch\"{o}nhammer,
{\it{Funcational renormalization group approach to correlated fermion systems}},
Rev. Mod. Phys. {\bf{84}}, 299 (2012).
%
\bibitem{Dupuis21}
N. Dupuis, L. Canet, A. Eichhorn, W. Metzner, J. M. Pawlowski, M. Tissier, and
N. Wschebor, {\it{The nonperturbative functional renormalization group and its applications}}, Phys. Rep. {\bf{910}}, 1 (2021).
%
\bibitem{Tarasevych21}
D. Tarasevych and P. Kopietz, {\it{Dissipative spin dynamics in hot quantum paramagnets}}, Phys. Rev. B {\bf{104}}, 024423 (2021). 
%
\bibitem{Krieg19}
J. Krieg and P. Kopietz, {\it{Exact renormalization group for quantum spin systems}},
Phys. Rev. B {\bf{99}}, 060403(R)  (2019).
%
\bibitem{Kawasaki66}
K. Kawasaki, {\it{Correlation Function Approach to the Transport Coefficients near the Critical Point. I}},
Phys. Rev. {\bf{150}}, 291 (1966).
%
\bibitem{Pomeau75}
Y. Pomeau and P. Resibois, {\it{Time dependent correlation functions and mode-mode coupling theories}},  Phys. Rep, {\bf{19}}, 64 (1975).
%
\bibitem{Goetze99}
W. G\"{o}tze, {\it{Recent tests of the mode-coupling theory for glassy dynamics}}, J. Phys.: Condens. Matter {\bf{11}}, A1 (1999).
%
\bibitem{Das04}
S. P. Das, {\it{Mode-coupling theory and the glas transition in supercooled liquids}},
Rev. Mod. Phys. {\bf{76}}, 785 (2004).
%
\bibitem{Mori65}
H. Mori, {\it{Transport, Collective Motion, and Brownian Motion}},
Prog. Theor. Phys. {\bf{33}}, 423 (1965).
%
\bibitem{Holm93}
C. Holm and W. Janke, {\it{Critical exponents of the classical three-dimensional
Heisenberg model: A single-cluster Monte Carlo study}}, Phys.Rev. B {\bf{48}}, 936 (1993).
%
\bibitem{footnoteCor}
Such non-analytic corrections to diffusion appear also in the high temperature limit $T \gg J$, which was previously discussed by us in Ref.~[\onlinecite{Tarasevych21}].
In our comparison with experiments measuring the coefficient 
$C_3$ defined via Eq.~(\ref{eq:C3def}) at high temperatures we have taken only
the contribution from the diffusion pole into account, which amounts to using a 
Lorentzian approximation $\Delta(\bm{k},i\omega)\approx \Delta(\bm{k},0)$ in 
our solution for $S(\bm{k},\omega)$.
%
\bibitem{Timusk99}
T. Timusk and B. Statt, {\it{The pseudogap in high-temperature superconductors: an experimental survey}}, Rep. Prog. Phys. {\bf{62}}, 61 (1999).
%
\bibitem{Takahashi86}
M. Takahashi, {\it{Quantum Heisenberg Ferromagnets in One and Two Dimensions at Low Temperature}}, Prog. Theor. Phys. Suppl. {\bf{87}}, 233 (1986).
%
\bibitem{Takahashi87}
M. Takahashi, {\it{Few-Dimensional Heisenberg Ferromagnets at Low Temperature}}, Phys. Rev. Lett. {\bf{58}}, 318 (1987).
%
\bibitem{Kopietz89}
P. Kopietz, {\it{Low-temperature behavior of the correlation length and the susceptibility of the
ferromagnetic quantum Heisenberg chain}}, Phys. Rev. B {\bf{40}}, 5194 (1989).
%
\bibitem{Takahashi90}
M. Takahashi, {\it{Dynamics of Heisenberg ferromagnets at low temperature}}, Phys. Rev. {\bf{B}} 42, 766 (1990).
%
\bibitem{Reiter93}
G. Reiter, {\it{Comment on ``Dynamics of Heisenberg ferromagnets at low temperature''}},
Phys. Rev. B {\bf{47}}, 8335 (1993).
%
\bibitem{Takahashi93}
M. Takahashi, {\it{Reply to ``Comment on `Dynamics of Heisenberg ferromagnets at low temperature' ''}},  Phys. Rev. B {\bf{47}}, 8336 (1993).
%
\bibitem{Karbach97}
M. Karbach,  G. M\"{u}ller, A. H. Bougourzi, A. Fledderjohann, and K.-H. M\"{u}tter,
{\it{Two-spinon dynamic structure factor of the one-dimensional s = 1/2
Heisenberg antiferromagnet}}, 
 Phys. Rev. B {\bf{55}}, 12510 (1997).
%
\bibitem{Caux08}
J.-S. Caux and R. Hagemans, {\it{The four-spinon dynamical structure factor
of the Heisenberg chain}}, J. Stat. Mech.: Theory Exp. P12013 (2006).
%
\bibitem{Mourigal13}
M. Mourigal, M. Enderle, A. Kl\"{o}pperpieper, J.-S. Caux, A. Stunault, and 
H. M. R{\o}nnow,
{\it{Fractional spinon excitations in the quantum Heisenberg antiferromagnetic chain}}, Nature Phys. {\bf{9}}, 413 (2013).
%
\bibitem{Dupont20}
M. Dupont and J. E. Moore, {\it{Universal spin dynamics in infinite-temperature one-dimensional quantum magnets}},
Phys. Rev. B {\bf{101}},  121106(R) (2020).
%
\bibitem{Bulchandani21}
V. B. Bulchandani, S. Gopalakrishnan, and E. Ilievski, {\it{Superdiffusion in spin chains}},
J. Stat. Mech. 084001 (2021).
%
\bibitem{Kopietz89b}
P. Kopietz and S. Chakravarty, {\it{Low-temperature behavior of the correlation length and the susceptibility
of a quantum Heisenberg ferromagnet in two dimensions}}, Phys. Rev. B {\bf{40}}, 4858 (1989).
%
\bibitem{Goll19}
R. Goll, D. Tarasevych, J. Krieg, and P. Kopietz,
{\it{Spin functional renormalization group for quantum Heisenberg ferromagnets:Magnetization and magnon damping in two dimensions}}, Phys. Rev. B {\bf{100}}, 174424 (2019).
%
\bibitem{Goll20}
R. Goll, A. R\"{u}ckriegel, and P. Kopietz, {\it{Zero-magnon sound in quantum Heisenberg ferromagnets}}, Phys. Rev. B {\bf{102}}, 224437 (2020).
%
\bibitem{Vaks68}
V. G. Vaks, A. I. Larkin, and S. A. Pikin, {\it{Thermodynamics of an ideal ferromagnetic substance}},
Zh. Eksp. Teor. Fiz. {\bf{53}}, 281 (1967) [Sov. Phys. JETP {\bf{26}}, 188 (1968)].
%
\bibitem{Vaks68b}
V. G. Vaks, A. I. Larkin, and S. A. Pikin,
{\it{Spin waves and correlation functions in a ferromagnetic}},
Zh. Eksp. Teor. Fiz. {\bf{53}}, 1089 (1967) [Sov. Phys. JETP {\bf{26}}, 647 (1968)].
%
\bibitem{Alessio16}
L. D'Alessio, Y. Kafri, A. Polkovnikov, and M. Rigol,
{\it{From quantum chaos and eigenstate thermalization to statistical mechanics and
thermodynamics}}, Adv. Phys. {\bf{65}}, 239 (2016).
%
\bibitem{Chiba20}
Y. Chiba, K. Asano, and A. Shimizu,
{\it{Anomalous behavior of Magnetic Susceptibility by Quench Experiments in Isolated Quantum Systems}}, Phys. Rev. Lett. {\bf{124}}, 110609 (2020).
%
%
%
%
%
%



\end{thebibliography}
\end{document}